# Rate Dependency in Steady-State Upscaling

**Lars Hov Odsæter**[1] · **Carl Fredrik Berg**[2] · **Alf Birger Rustad**[2]

**Abstract** Steady-state upscaling of relative permeability is studied for a range of reservoir models. Both rate-dependent upscaling and upscaling in the capillary and viscous limits are considered. In particular, we study fluvial depositional systems, which represent a large and important class of reservoirs. Numerical examples show that steady-state upscaling is rate dependent, in accordance with previous work. In this respect we introduce a scale-dependent capillary number to estimate the balance between viscous and capillary forces. The difference between the limit solutions can be large, and we show that the intermediate flow rates can span several orders of magnitude. This substantiate the need for rate-dependent steady-state upscaling in a range of flow scenarios. We demonstrate that steady-state upscaling converges from the capillary to the viscous limit solution as the flow rate increases, and we identify a simple synthetic model where the convergence fails to be monotone. Two different sets of boundary conditions were tested, but had only minor effects on the presented reservoir models. Finally, we demonstrate the applicability of steady-state upscaling by performing dynamic flow simulation at the reservoir scale, both on fine-scaled and on upscaled models. The considered model is viscous dominated for realistic flow rates, and the simulation results indicate that viscous limit upscaling is appropriate.

✉ Lars Hov Odsæter
lars.odsater@math.ntnu.no

Carl Fredrik Berg
carlpaatur@hotmail.com

Alf Birger Rustad
abir@statoil.com

1 Department of Mathematical Sciences, NTNU Norwegian University of Science and Technology, Alfred Getz' vei 1, 7491 Trondheim, Norway

2 Statoil R&D Center, Arkitekt Ebbels veg 10, Rotvoll, 7053 Trondheim, Norway

## 1 Introduction

Hydrocarbon reservoirs are generally heterogeneous on different length scales (Kløv et al. 2003; Lerdahl et al. 2005; Aarnes et al. 2007). Due to computational limitations, it is not always possible to include important heterogeneities into a full field reservoir simulation. Nevertheless, small-scale heterogeneities may be important for the global flow and therefore should be taken into account. Upscaling is a well-known technique in this respect, see, e.g. Durlofsky (2005) for an overview of different upscaling methods. The overall goal for all methods is to replace a fine-scale model with a coarse model by producing effective properties for each coarse cell. For upscaling of nonbulk flow properties like permeability and relative permeability, one may divide upscaling into two main classes. These are averaging methods and flow-based methods. The latter are regarded to be more accurate, but they are also computationally more demanding as they require solutions to flow equations. Furthermore, one can roughly divide the flow-based methods into local and global methods. Global methods, see, e.g. Kyte and Berry (1975) and Zhang et al. (2006), require solutions to the full fine-scale model, while local methods only consider the fine-scale region corresponding to the target coarse cell. Noting that this is a simplified view, there are also combinations of the two. Global methods typically address the common task of upscaling from a geo-cellular model to a simulation model. Local methods are more generic in nature and are convenient for determining effective flow properties for rock types or facies in geo-cellular models, see, e.g. Pickup and Stephen (2000), Kløv et al. (2003), Rustad et al. (2008) and Nordahl et al. (2014).

The work presented here can be seen as a continuation of the work in Nordahl et al., with a focus on multiphase flow parameters. More precisely, we investigate the effects of sedimentary heterogeneity on oil–water relative permeability. Our approach is based on steady-state upscaling, a local upscaling method which has been subject to extensive research over the last 20 years, see, e.g. Smith (1991), Ekrann et al. (1996), Kumar et al. (1997), Ekrann and Aasen (2000), Pickup et al. (2000), Lohne et al. (2006) and Jonoud and Jackson (2008). As most literature, we consider incompressible, immiscible two-phase flow and neglect gravity. From the geological perspective we focus on fluvial depositional systems. These are particularly interesting from an upscaling point of view because of their heterogeneity structure. Fluvial systems typically consist of channels, crevasses and background. The channels are typically highly permeable, with an internal depth trend. Crevasses typically have lower permeability, but still contribute significantly to flow and volume. The background material is typically tight with little or no permeability. While channels can be around five meters tall (depending on the size of the depositional system), they can meander and erode into each other, stacking in complex patterns. Crevasses may be less than a meter tall. Hence, we have extreme contrast between reservoir properties at the scale of several meters, with a geometry we cannot represent in current full field simulation models. That is, the finest grid resolution we realistically can run full field flow simulations on is typically not able to represent the geometry of the main sedimentary heterogeneities of fluvial reservoirs. Moreover, fluvial reservoirs are among the most common depositional systems found in reservoirs.

For any type of flow-based upscaling method, the imposed boundary conditions can have an important effect on the upscaling results. This is one of the aspects that make upscaling ambiguous and a difficult exercise. Moreover, disregarding gravity, the viscous and capillary forces are the acting forces on the fluids in a reservoir. The results from steady-state upscaling is known to be rate dependent due to the balance of these two forces (Ekrann et al. 1996; Kumar et al. 1997; Ekrann and Aasen 2000; Virnovsky et al. 2004; Lohne et al. 2006). Capillary forces dominate for small flow rates or on small-scale heterogeneities, while viscous forces dominate for high flow rates or on larger scales. Hence, as the flow rate tends to zero, it is expected that the viscous forces can be neglected. Similarly, for high flow rates it is expected that capillary forces can be neglected. These limits are referred to as the capillary and viscous limits, respectively, and they are interesting since they are computationally less demanding. However, these limits are only encountered in practice at some scales and some parts of the reservoir.

The effect of small-scale heterogeneities on relative permeability, typically heterolithic reservoir zones, has been studied before (e.g., Rustad et al. (2008)). Two important publications where the transition between capillary and viscous forces are studied are found in Virnovsky et al. (2004) and Lohne et al. (2006). There is a couple of differences to note between the work presented here and in Virnovsky et al. (2004) and Lohne et al. (2006). The first is that the steady-state simulations done here are on three-dimensional models with corner-point grids. This enables the use of the ReservoirStudio software[1] providing realistic models of heterogeneity at a scale where the balance between capillary and viscous forces is unclear. Secondly, all upscaling in this work are performed with codes from the Open Porous Media project (OPM),[2] and are freely available under the GNU GPL license. However, models and geomodelling packages are still unfortunately proprietary and hence not easy to reproduce results from.

It would be valuable to be able to decide whenever the capillary or viscous limits are fair approximations for steady-state flow. A recent work (Jonoud and Jackson 2008) addresses this problem, but originally for two-dimensional models. Although it might be possible to extend their method to three dimensions, the ambiguity that is present for two-dimensional models is even more involved when moving to three dimensions. Our contribution to this problem is a scale-dependent capillary number, which is shown to better predict the balance between capillary and viscous forces in our models compared to the traditional capillary number. For the reservoirs considered herein, heterogeneity typically introduces large differences in capillary pressure compared to capillary pressure gradients from saturation distribution inside a single rock type. For this reason, we restrict our attention to models with several rock types.

This paper is outlined as follows. We start by presenting the governing flow equations in Sect. 2. Thereafter, in Sect. 3, we explain steady-state upscaling in more detail. In particular, we discuss and define appropriate boundary conditions for the steady-state flow equations. We also describe the capillary and viscus limit approaches. In Sect. 4 we introduce a dimensionless scale-dependent capillary number to represent the capillary and viscous force balance in our models. The aim is to be able to predict when steady-state upscaling differs from its capillary and viscous limits. In Sect. 5 we test the steady-state upscaling approaches on a variety of reservoir models, both synthetic and realistic. The overall goal of this section is to get a better understanding of steady-state upscaling and to address the main advantages and challenges. We show how the results depend on the flow rate and on the choice of boundary conditions. In the last part, we consider a larger fluvial reservoir model similar to those considered in Nordahl et al. (2014). The aim is to study how well steady-state upscaling is able to reproduce fine-scale flow pattern and production data in a dynamic flow scenario. Finally, in Sect. 6, we discuss the results of this work and make the final conclusions.

## 2 Flow Equations

In this section we present the governing equations for incompressible immiscible two-phase flow in porous media. Let oil ($o$) and water ($w$) denote the two phases $p$. The equation of

[1] ReservoirStudio$^{\text{TM}}$ is a proprietary software from Geomodeling, www.geomodeling.com.

[2] The Open Porous Media project, http://www.opm-project.org.

continuity reads

$$\phi \frac{\partial S_p}{\partial t} = -\nabla \cdot \mathbf{u}_p, \qquad p = o, w, \tag{1}$$

where $\phi = \phi(\mathbf{x})$ is the porosity at position $\mathbf{x} = \langle x, y, z \rangle$, while $S_p = S_p(\mathbf{x}, t)$ and $\mathbf{u}_p = \mathbf{u}_p(\mathbf{x}, t)$ are the saturation and velocity, respectively, at position $\mathbf{x}$ and time $t$. We assume Darcy flow, i.e.,

$$\mathbf{u}_p = -\frac{\mathsf{K} k_{rp}}{\mu_p} \nabla p_p, \qquad p = o, w, \tag{2}$$

where we have neglected gravity. Furthermore, $\mathsf{K} = \mathsf{K}(\mathbf{x})$ is the absolute permeability tensor, and $k_{rp} = k_{rp}(\mathbf{x}, S_w)$, $\mu_p$ and $p_p = p_p(\mathbf{x}, t)$ are the relative permeability, viscosity and pressure of phase $p$, respectively. If we further introduce the mobility of phase $p$,

$$\lambda_p = \lambda_p(\mathbf{x}, S_w) = \frac{k_{rp}(\mathbf{x}, S_w)}{\mu_p}, \tag{3}$$

and combine Eqs. (1) and (2), we get

$$\phi \frac{\partial S_p}{\partial t} = \nabla \cdot \left[\mathsf{K} \lambda_p \nabla p_p\right], \qquad p = o, w. \tag{4}$$

To complete the system of equations, we have that

$$S_w + S_o = 1, \qquad \text{and} \tag{5}$$

$$p_o - p_w = p_c, \tag{6}$$

where $p_c = p_c(\mathbf{x}, S_w)$ is the capillary pressure. In total we have four equations, Eqs. (4–6), and four unknowns, $S_p$ and $p_p$, for $p = o, w$.

We follow the approach of Chavent and Jaffre (1986) and introduce a global pressure $p$, defined as

$$p = p_o - \hat{p}, \qquad \hat{p}(S_w) = \int_1^{S_w} f_w(\xi) \frac{\partial p_c}{\partial S_w}(\xi)\, \mathrm{d}\xi,$$

where $f_w(S_w) = \frac{\lambda_w(S_w)}{\lambda(S_w)}$ is the fractional flow of water and $\lambda = \lambda_w + \lambda_o$ is the total mobility. We can now rewrite the coupled system of Eqs. (4–6) as

$$\nabla \cdot \mathbf{u} = 0, \qquad \mathbf{u} = -\mathsf{K} \lambda \nabla p, \tag{7}$$

$$\phi \frac{\partial S_w}{\partial t} + \nabla \cdot \left[f_w(\mathbf{u} + \mathsf{K} \lambda_o \nabla p_c)\right] = 0. \tag{8}$$

The pressure equation (Eq. (7)) is elliptic, while the transport equation (Eq. (8)) is parabolic or hyperbolic dependent on the ratio between the terms. The equations are nonlinearly coupled due to the pressure dependency in $u$ and the saturation dependency in $\lambda_p$ and $p_c$.

When deriving the pressure equation, it is assumed that the capillary pressure $p_c$ is a monotone function of $S_w$ only, thus independent on the spatial position. In general, this assumption is not satisfied, but we believe that the introduced error is small. Testing and comparison with a fully implicit solver substantiate this.

At steady state, $\frac{\partial S_p}{\partial t} = 0$, so Eq. (4) reduces to

$$\nabla \cdot \left[\mathsf{K} \lambda_p \nabla p_p\right] = 0. \tag{9}$$

Furthermore, the system (7, 8) can at steady state be written as

$$\mathbf{u} \cdot \nabla f_w + \nabla \cdot (\mathsf{K} \lambda_o f_w \nabla p_c) = 0. \tag{10}$$

## 3 Steady-State Upscaling

Steady-state upscaling, see, e.g. Ekrann and Aasen (2000), is a local upscaling method. Given a target coarse cell with fine-scale heterogeneity, we seek to find an effective (upscaled) relative permeability function as if the cell was homogeneous. Steady-state upscaling can be divided into three main steps:

1. Calculate the fine-scale saturation distribution, $S_w(\mathbf{x})$, at steady state.
2. Calculate the phase permeability distribution, $\mathsf{K}_p(\mathbf{x}, S_w) = k_{rp}(\mathbf{x}, S_w)\mathsf{K}(\mathbf{x})$, from input data.
3. Perform single-phase upscaling on each phase separately to calculate the upscaled phase permeability tensor, $\tilde{\mathsf{K}}_p$.

In Step 3, single-phase upscaling corresponds to solving the steady-state equation (Eq. (9)) for each phase and then let $\tilde{\mathsf{K}}_p$ be the phase permeability tensor which preserves the flux over each boundary face. The upscaled relative permeability tensor is now given as $\tilde{\mathsf{K}}_p \cdot \tilde{\mathsf{K}}^{-1}$, where $\tilde{\mathsf{K}}$ is the upscaled (absolute) permeability tensor. The procedure above gives an upscaled relative permeability tensor valid for the upscaled water saturation, $\tilde{S}_w$, which is simply the volume weighted average. By varying the initial saturation and the fractional flow at inlet, one produces a sequence of upscaled tensors for different upscaled saturation points, giving an upscaled relative permeability curve. In general, Step 1 is the most demanding, as it requires a full flow simulation to reach the steady-state distribution, $S_w(\mathbf{x})$. However, in the capillary and viscous limits, $S_w(\mathbf{x})$ can be calculated directly from the input data.

### 3.1 General Steady-State Upscaling

The domain $\Omega$ which is to be upscaled for, is assumed to be formed like a shoe-box, that is, a regular hexahedron. This is a natural assumption, as $\Omega$ usually refers to a coarse simulation cell. Let $\partial\Omega^{\zeta,i}$ for $\zeta = x, y, z$ and $i = 1, 2$ denote the six boundary faces on $\Omega$ as illustrated in Fig. 1. To obtain the steady-state saturation distribution, we solve Eqs. (7) and (8) over the domain $\Omega$ until steady state is reached, i.e., until the saturation distribution no longer changes with time within a given tolerance.

The flow equations must be accompanied with some appropriate boundary conditions (BCs). In this work we use two sets of BCs, denoted *fixed* and *periodic*. To induce flow, a pressure drop $\Delta p$ is enforced in one of the Cartesian directions, hereafter denoted the pressure drop direction. For fixed BCs we set $p = \Delta p$ and $f_w = g(\mathbf{x})$ on inlet, and $p = 0$ on outlet, where $g(\mathbf{x})$ is a known function. For periodic BCs we set the flux and the fractional flow to be periodic on inlet/outlet. This implies that what flows out at outlet flows in at inlet. A main distinction from fixed BCs is that $f_w$ and $p$ are allowed to vary across $\partial\Omega$. Furthermore, we set the pressure difference between inlet and outlet to $\Delta p$. For both periodic and fixed BCs we use no-flow conditions on all other boundaries. Notice that this differs from the common notion of periodic BCs, used, e.g. in Durlofsky (2005), as we do not induce periodic flow on boundaries parallel to the pressure drop direction. With this simplification, we neglect cross flow and our upscaled tensor will be diagonal. For reservoir models where the main heterogeneity structure is aligned with the pressure drop direction, this is a reasonable assumption.

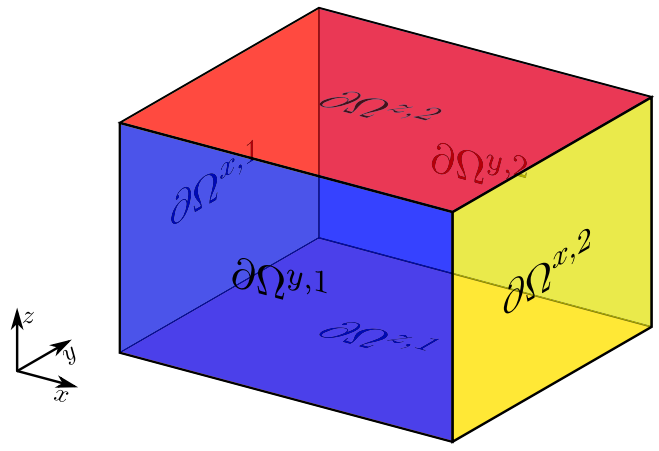

| *Fixed* | *Periodic* |
|---|---|
| $p\|_{\partial\Omega^{\eta,1}} = \Delta p,$ | $p\|_{\partial\Omega^{\eta,1}} = p\|_{\partial\Omega^{\eta,2}} + \Delta p,$ |
| $p\|_{\partial\Omega^{\eta,2}} = 0,$ | $(\mathbf{u}\cdot\mathbf{n})\|_{\partial\Omega^{\xi,i}} = 0,$ |
| $(\mathbf{u}\cdot\mathbf{n})\|_{\partial\Omega^{\xi,i}} = 0,$ | $(\mathbf{u}\cdot\mathbf{n})\|_{\partial\Omega^{\eta,1}} = -(\mathbf{u}\cdot\mathbf{n})\|_{\partial\Omega^{\eta,2}},$ |
| $f_w\|_{\partial\Omega^{\eta,1}} = g(x).$ | $f_w\|_{\partial\Omega^{\eta,1}} = f_w\|_{\partial\Omega^{\eta,2}}.$ |

**Fig. 1** Boundary conditions for the two-phase incompressible immiscible problem, Eqs. (7) and (8). To the *left* we see an illustration of the domain $\Omega$ which is to upscaled for, along with the naming conventions for the six boundary faces. To the *right*, precise formulations of the fixed and periodic BCs are given, where $\eta$ denotes the pressure drop direction, and $\xi$ denotes the two directions perpendicular to $\eta$

The two sets of BCs are given explicitly in Fig. 1. We let $\mathbf{n}$ denote the outward pointing unit normal vector, $\eta$ the pressure drop direction, $\xi$ the two other directions, and $f|_{\partial\Omega}$ the function $f$ restricted to $\partial\Omega$. Additionally, for periodic BCs, it is necessary to specify the global pressure at one point in the domain for the system to be well-posed. Furthermore, an initial guess of the saturation distribution, $S^0(\mathbf{x})$, must be specified. For the periodic BCs, we use a face-to-face connection on the periodic boundaries, and we have assumed that the grids on the inlet and outlet faces match each other. This assumption is not satisfied for all realistic reservoir models, but may be overcome by mirroring the model in the pressure drop direction. Notice that with periodic BCs no fluids are allowed to enter or leave the model. Hence, the upscaled saturation at steady state is equal to the initial average saturation.

### 3.2 Capillary Limit

In the capillary limit, see, e.g. Ekrann et al. (1996), we neglect gravity and viscous forces. In this limit the fluid velocity vanishes, and hence Eq. (10) reduces to

$$\nabla \cdot (\mathsf{K}\lambda_o f_w \nabla p_c) = 0.$$

Since this should hold for the whole domain $\Omega$ and since the product $\mathsf{K}\lambda_o f_w$ is spatially dependent, we must require that $\nabla p_c = 0$, or equivalently that $p_c$ is constant throughout the model. Thus, at a given (constant) capillary pressure, $P_c$, the fine-scale saturation distribution is determined by solving

$$p_c(\mathbf{x}, S_w) = P_c \tag{11}$$

for $S_w$. Since $p_c(\mathbf{x}, S_w)$ is a strictly monotonically decreasing function with respect to $S_w$, this has a unique solution. Observe that this approach also produces an upscaled capillary pressure function,

$$\tilde{p}_c(\tilde{S}_w) = P_c. \tag{12}$$

In the following we refer to the capillary limit solution by CL.

### 3.3 Viscous Limit

In the viscous limit, see, e.g. Ekrann and Aasen (2000), we neglect gravity and capillary forces. Thus, Eq. (10) reduces to

$$\mathbf{u} \cdot \nabla f_w = 0. \tag{13}$$

This means that the fractional flow is constant along streamlines. The viscous limit solution is therefore determined by the fractional flow on the inlet boundary and is thus not unique. In general, if we know the fractional flow of water throughout the model, $F_w(\mathbf{x})$, we can calculate the saturation distribution by solving

$$f_w(\mathbf{x}, S_w) = F_w(\mathbf{x}) \tag{14}$$

for $S_w$. Under the standard assumptions that $k_{rp} \geq 0$ for $p = o, w$, and that $k_{rw}$ and $k_{ro}$ are strictly monotonically increasing and decreasing functions with respect to $S_w$, respectively, it can be shown that $f_w(\mathbf{x}, S_w)$ is a strictly monotonically increasing function with respect to $S_w$. Thus, Eq. (14) has a unique solution.

Calculating the streamlines for a two-phase three-dimensional problem is in general non-trivial. A class of problems for which the streamlines are identical to their single-phase counterparts has been identified (Ekrann and Aasen 2000). This class is characterized by $\mathbf{u} \cdot \nabla\lambda = 0$, which holds if the relative permeability curves as functions of normalized water saturation are independent on $\mathbf{x}$. For general problems, tracking streamlines is dependent on the discrete model, so the only steady-state viscous limit we can obtain is the one with constant fractional flow throughout the model. This is equivalent to assuming constant fractional flow on the inlet boundary. In the following, when we refer to VL we mean the viscous limit solution with constant fractional flow.

Based on the discussion above, it is necessary to have constant fractional flow on the inlet boundary if we want the general steady-state approach to converge to VL for high flow rates. For fixed BCs this is achieved by setting $g(\mathbf{x})$ equal to the constant fractional flow that we want to upscale for, i.e., $g(\mathbf{x}) \equiv F_w$ cf. Figure 1. For periodic BCs the fractional flow does not necessarily converge toward a constant for high flow rates in an analytical sense. In our discretization of the fluid flow, we assume a full mixing of fluids in the grid cells. We therefore avoid mixing only when streamlines are parallel to cell boundaries. Thus, for most models, as demonstrated by numerical experiments in Sect. 5.2, we obtain constant fractional flow for high pressure drops. A layered model, as we consider in Sect. 5.1, is an example where streamlines are parallel to cell boundaries, and where the fractional flow is not necessarily constant for high pressure drops.

In contrast to the CL approach, there is no natural way of upscaling capillary pressure in the VL approach. Hence, we neglect capillary forces by letting $p_c \equiv 0$ when considering viscous limit upscaled simulation models in Sect. 5.

## 4 Scale-dependent Capillary Number

Upscaling of two-phase flow is dependent on the balance between viscous and capillary forces. Evaluation of the force balance helps to determine which upscaling method is most appropriate. In a multiscale upscaling process, disregarding viscous forces is typically a fair approximation for models on the smallest scale (e.g., lamina scale, mm to cm), while in a full field reservoir model the impact from capillary forces is often negligible. The balance between viscous and capillary forces is thus scale dependent.

At the microscopic (interstitial) scale, the viscous–capillary force balance is traditionally described by a microscopic capillary number. There exists different forms of the capillary number, see, e.g., (Lake 1989, Tables 2, 3). Most are similar to

$$\mathrm{Ca} = \frac{\mu u}{k_r \sigma \cos(\theta)} \simeq \frac{\mu u}{\sigma}, \tag{15}$$

where $u$ is Darcy velocity, $\sigma$ is the interfacial tension, and $\theta$ is the contact angle, often set to $0^{\circ}$. The relative permeability, $k_r$, is often set to 1 (Dullien 1992). Observe that the capillary number, Ca, increases with increasing fluid velocity. This reflects that on the microscopic scale the capillary forces dominate for small flow rates, while viscous forces dominate for high flow rates. This transition is reflected in capillary desaturation curves, describing the relationship between residual oil saturation and the capillary number. Such capillary desaturation curves show that below a critical capillary number the residual oil saturation remains constant, while the residual oil saturation decreases with increasing capillary numbers above the critical number (Morrow et al. 1988; Dullien 1992).

In this paper we deal with fluids and flow rates for which the capillary number is below the critical value. Thus, at the pore scale the residual oil saturation is assumed constant, and the fluid distribution will be dominated by capillary forces. Instead we are interested in the balance of viscous and capillary forces on a macroscopic (Darcy) scale. For models representing reservoir heterogeneities, individual rock types are populated with particular capillary pressure curves. Such different capillary pressure curves introduce capillary forces not represented by the traditional capillary numbers.

We seek good static approximations of the magnitudes of the viscous and capillary forces that take different capillary pressure curves into account. A characteristic magnitude of the viscous forces is given by $|\Delta p|/L$, where $L$ is the length of the model in the direction of the pressure drop, $\Delta p$. Note that this is a static parameter for typical upscaling procedures. As the gradient of the capillary pressure is dependent on the saturation distribution, a dynamic variable, any static characteristic for the capillary forces will be an approximation. We approximate the capillary forces with an estimate of the norm of the capillary pressure gradient,

$$\mathcal{C} = \frac{\sum_{ij}\left((V_i\phi_i + V_j\phi_j)\Delta p_{c,ij}/l_{ij}\right)}{\sum_{ij}(V_i\phi_i + V_j\phi_j)} \approx \|\nabla p_c\|\,, \tag{16}$$

where the sum is over neighboring cells $ij$ with length $l_{ij}$ between the cell centers, $V_i$ is the volume of cell $i$, $\phi_i$ is the porosity of cell $i$, and $\Delta p_{c,ij}$ is a measure of the absolute difference in capillary pressure between the neighboring cells. Hence, $\mathcal{C}$ is the pore volume weighted average of the estimated capillary pressure gradient, $\Delta p_{c,ij}/l_{ij}$, over cell interfaces. In this work we define $\Delta p_{c,ij}$ as an integral average, such that

$$\Delta p_{c,ij} = \frac{1}{1-\tilde{S}_{orw}-\tilde{S}_{wir}-2\delta}\int_{\tilde{S}_{wir}+\delta}^{1-\tilde{S}_{orw}-\delta}\left|p_{c,i}(S_w^{VL}) - p_{c,j}(S_w^{VL})\right|\,\mathrm{d}\tilde{S}_w^{VL},$$

where $S_w^{VL}$ is to be interpreted as the water saturation distribution at VL with upscaled water saturation equal to $\tilde{S}_w^{VL}$. Furthermore, $p_{c,i}(S_w^{VL})$ is the capillary pressure in cell $i$ for this saturation distribution. The endpoint saturations, $\tilde{S}_{wir}$ and $\tilde{S}_{orw}$, refer to the upscaled irreducible water saturation and the upscaled residual oil saturation, respectively. The capillary pressure curves are typically very steep close to the endpoints. To obtain representative capillary pressure values, we neglect these regions when taking the integral average. We choose $\delta$ to be

$$\delta = \epsilon(\tilde{S}_{wor} - \tilde{S}_{wir}),$$

and set $\epsilon = 0.1$. We report on the stability of this choice in Sect. 5.2.

A macroscopic scale-dependent capillary number, $\mathcal{N}$, is now defined as the fraction between viscous and capillary forces,

$$\mathcal{N} = \frac{|\Delta p|/L}{\mathcal{C}} \approx \frac{\|\nabla p\|}{\|\nabla p_c\|}. \tag{17}$$

By using Eq. (16) as an approximation for the gradient of the capillary pressure, we neglect contributions from saturation gradients, e.g. inside a single rock type. For the fluvial reservoir type considered in this work, it is a fair assumption that differences in capillary pressure due to heterogeneity dominate the contributions from saturation gradients, especially close to a steady-state scenario.

The variation in capillary pressures in a small-scaled model is typically at the same order as the variation in a coarser model. Hence, $\mathcal{C}$ is typically smaller in a coarser model through the dependency on $l_{ij}$. The viscous pressure gradient, $\Delta p/L$, is on the other hand not affected by the grid size. Therefore, $\mathcal{N}$ typically decreases with model size, reflecting that capillary forces dominate on fine-scale models, while viscous forces dominate in coarser models.

## 5 Numerical Examples

In this section we apply the steady-state upscaling methods described in Sect. 3 on a range of models, both synthetic and realistic. The purpose is to study the rate dependency in steady-state upscaling and also the impact of different boundary conditions (BCs). In the last part of this section we consider a larger representative model and test the different upscaling regimes on reservoir simulation. The aim is to see how well steady-state upscaling is able to reproduce dynamic flow on reservoir scale.

For general steady-state upscaling, the pressure and saturation equations (Eq. (7, 8)) are solved sequentially. For the elliptic pressure equation, a mimetic finite difference method (Brezzi et al. 2005) is used, while an implicit Euler method with upstream weighting of the fractional flow is used for the saturation equation. The BCs for the flow-based local single-phase upscaling correspond to the ones used to solve the two-phase equations, i.e., as seen in Fig. 1, but without the fractional flow conditions. As the initial guess we use the capillary limit distribution unless a steady-state solution to a similar problem exists. All models are represented in corner-point grids. Fluid data are listed in Table 1.

### 5.1 Synthetic Models

In this section we consider simple synthetic models built up of two homogeneous isotropic rocks: one high permeability rock with $k = 1\,\mathrm{D}$ (rock 1), and one low permeability rock with $k = 1\,\mathrm{mD}$ (rock 2). The porosity is 0.1 for both rocks, and we use the same relative permeability curves and Leverett J-function curve,

$$k_{rw}(S_w) = S_w^2, \quad k_{ro}(S_w) = (1 - S_w)^2, \quad J(S_w) = \frac{1}{S_w} - \frac{1}{1 - S_w}. \tag{18}$$

The capillary pressure and the Leverett J-function are related as $p_c(S_w) = \sqrt{\phi/k}\sigma J(S_w)$, assuming a contact angle of 0°. Hence, the capillary pressure curves are different in the two rocks. As the relative permeability curves are equal for oil and water, only mirrored around $S_w = 0.5$, we only present upscaled water curves for better readability. The results and conclusions are the same for oil curves.

**Table 1** Fluid data used for all models in this work

| Parameter | Symbol | Value |
|---|---|---|
| Interfacial tension | $\sigma$ | 11 dynes/cm = 0.011 N/m |
| Viscosity, oil | $\mu_o$ | 1.78 cP = 0.00178 Pa s |
| Viscosity, water | $\mu_w$ | 0.33 cP = 0.00033 Pa s |

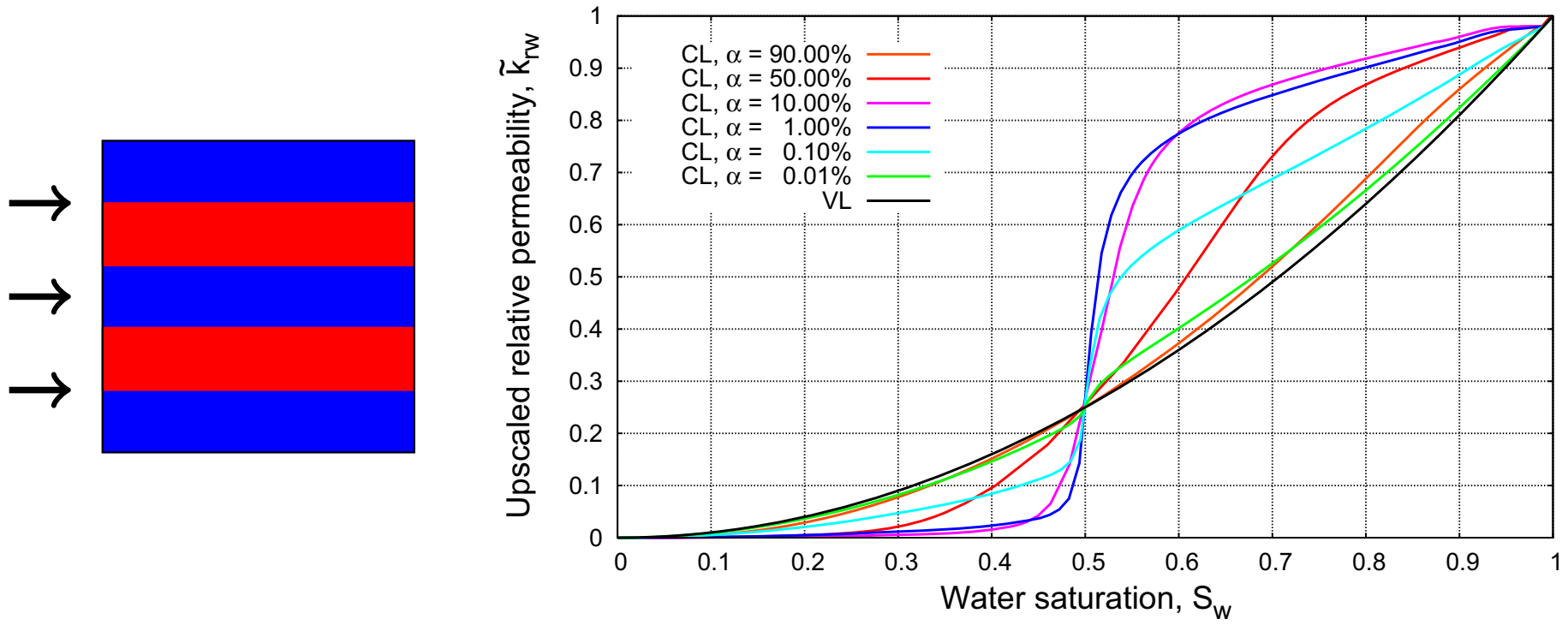

**(a)** Arithmetic averaging (flow aligned with the layers)

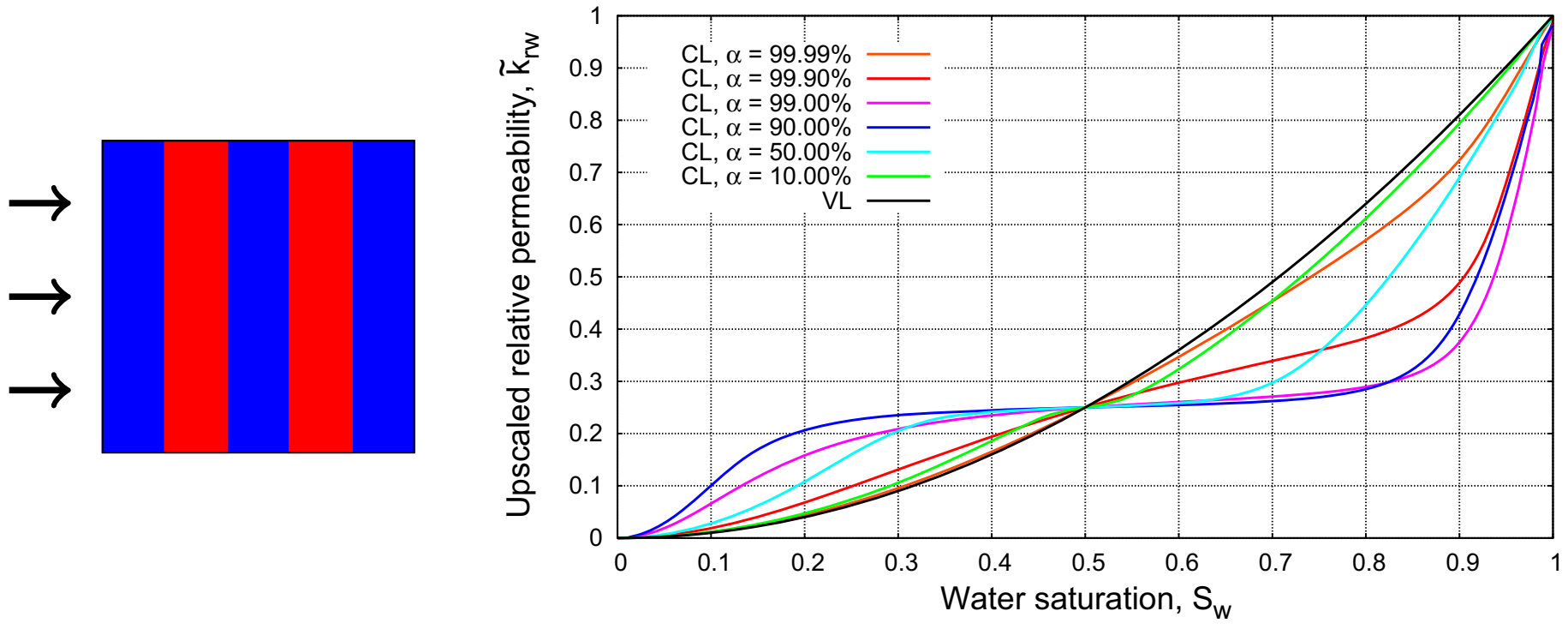

**(b)** Harmonic averaging (flow normal to the layers)

**Fig. 2** Analytic solutions for parallel models, where (**a**) flow is aligned with the layers, and (**b**) flow is normal to the layers. The volume fraction between rock 1 ($k = 1$ D) and rock 2 ($k = 1$ mD) is denoted $\alpha$. We only show water curves, as oil curves are equal, only mirrored around $S_w = 0.5$

First, we consider models where the two rock types alternate in parallel layers. If the flow is aligned with the layers or if the flow is normal to the layers, we can use arithmetic or harmonic volume weighted averaging, respectively. Let $\alpha$ denote the volume fraction between rock 1 and rock 2. Since the relative permeability curves are identical, the VL solution is independent on $\alpha$ and is equal to the input relative permeability curve. However, due to the difference in capillary pressure, the CL solution depends on $\alpha$.

In Fig. 2 the analytic results for different volume fractions are displayed. For $\alpha$ close to 0 or 100 %, the CL curves are close to the VL curves in both cases. When flow is aligned with the layers (Fig. 2a), the difference between CL and VL upscaling becomes significant at very small portions of the high permeability rock. For flow normal to the layers (Fig. 2b), the difference becomes significant for very small portions of the low permeability rock. This illustrates that even very thin high permeability channels or low permeability barriers can have large effects on the upscaling results. These results have also been used to verify our numerical upscaling routines.

Next, we apply steady-state upscaling on the three synthetic models depicted in Fig. 3. The rock and fluid data are all the same as in the analytic examples. All models are 100 m long

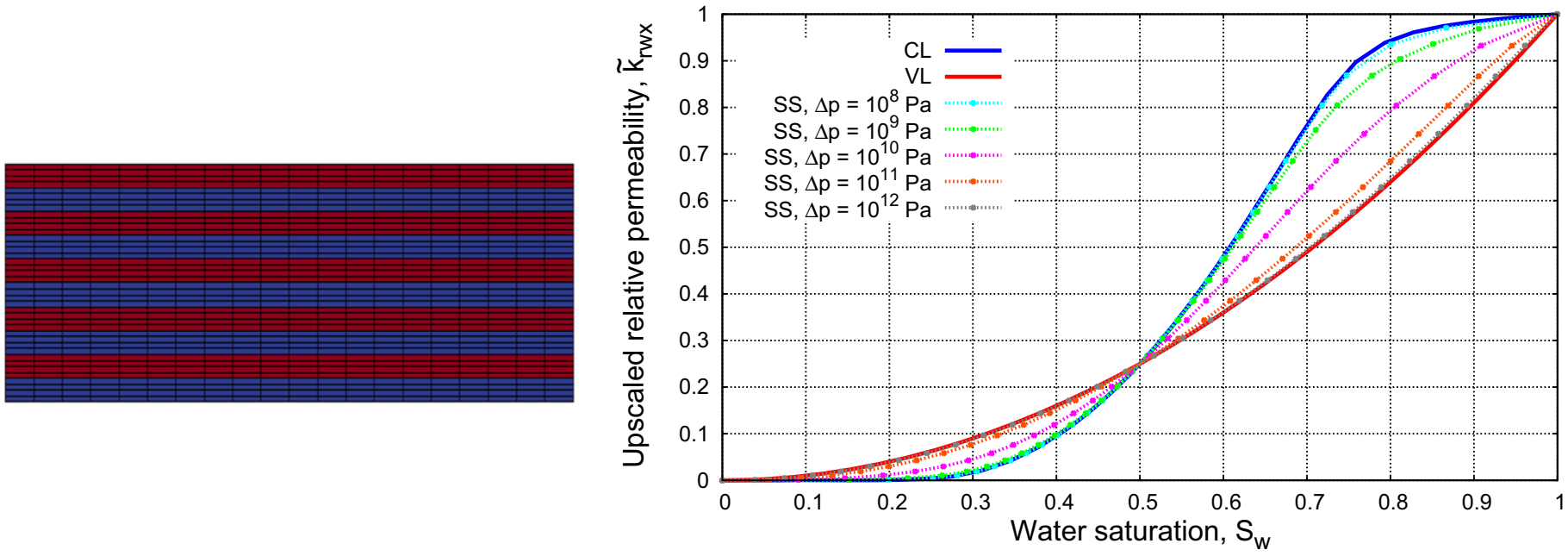

**(a)** Flow aligned with the layers

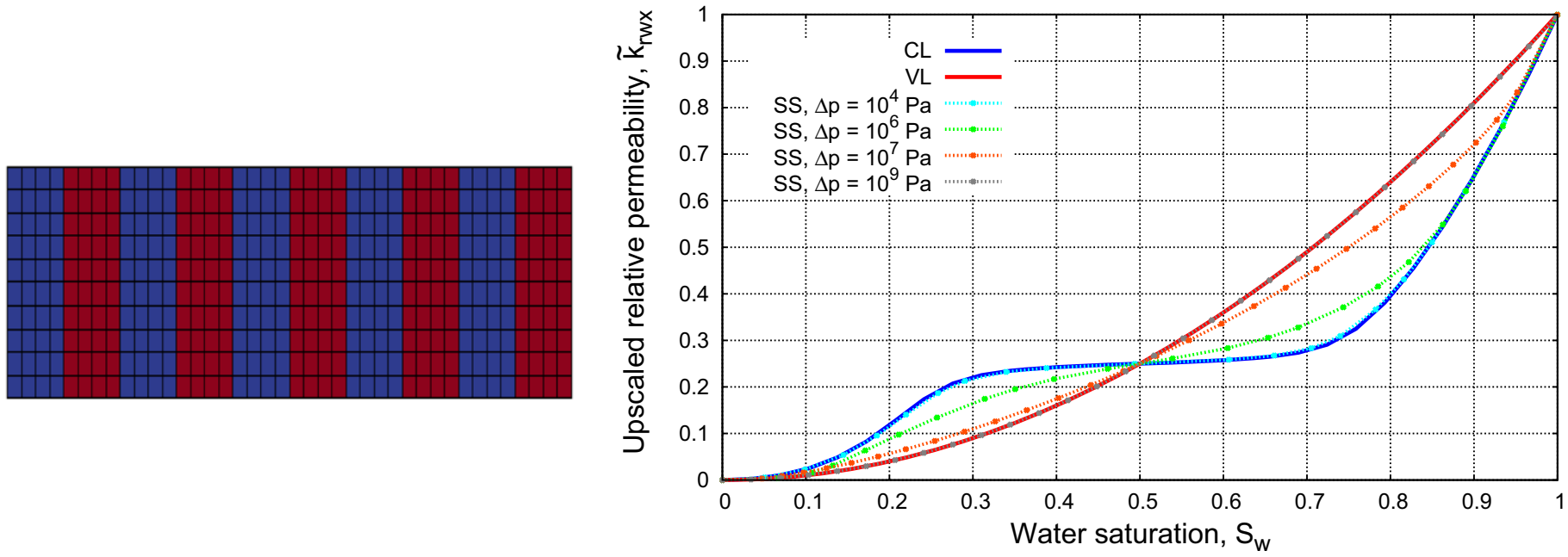

**(b)** Flow normal to the layers

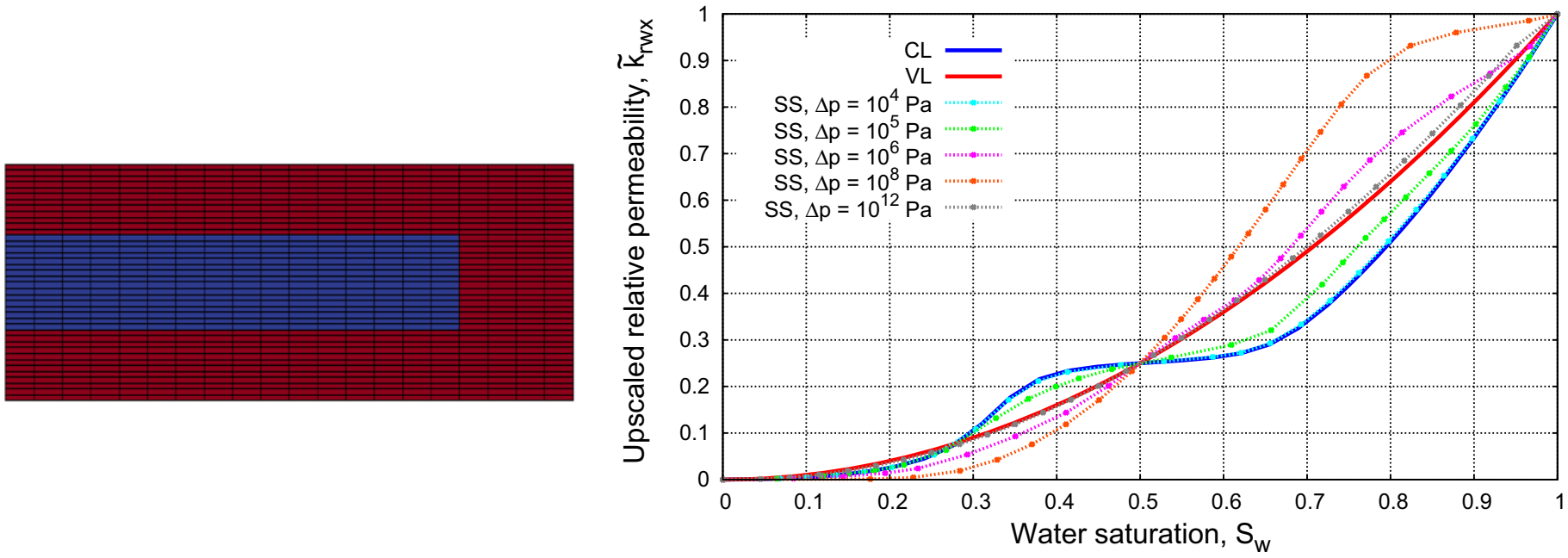

**(c)** Flow through a barrier

**Fig. 3** Synthetic test models. All models are $100 \times 5$ m. Rock 1 (*blue*) has $k = 1$ D and rock 2 (*red*) has $k = 1$ mD. Results for different pressure drops and the VL and CL solutions are shown to the right. The upscaled oil curves are equal, just mirrored around $S_w = 0.5$. The pressure drop direction is from left to right and fixed BCs are used

and 5 m thick. We consider flow in the $x$-direction, that is left to right. The results with fixed BCs for different pressure drops and in CL and VL are shown in Fig. 3. Both the microscopic capillary number, Ca, and the proposed macroscopic capillary number, $\mathcal{N}$, are presented in Table 2.

For all models, we observe that general steady-state upscaling converges toward CL as $\Delta p$ decreases, and toward VL as $\Delta p$ increases. This is as expected, since for small pressure drops capillary forces manage to redistribute the fluids toward the CL, while for large enough pressure drops, the viscous forces dominate. For the two first models (Fig. 3a, b), the convergence from CL to VL is monotonic and the general steady-state upscaling curves are bounded by the CL and VL curves. This property does not hold for the last model (Fig. 3c), where the high permeability rock is aligned with the flow, but not percolating.

**Table 2** The microscopic capillary number, Ca (Eq. (15)), interpreted as $\mathrm{Ca} = \frac{\mu u}{\sigma} = \frac{\tilde{k}\Delta p}{\sigma L}$, and the scale-dependent capillary number, $\mathcal{N}$ (Eq. (17)), as functions of the global pressure drop (in Pa), $\Delta p$, for the three models in Fig. 3

| Model | Ca | $\mathcal{N}$ |
|---|---|---|
| Figure 3a | $4.5 \cdot 10^{-13} \Delta p$ | $5.1 \cdot 10^{-8} \Delta p$ |
| Figure 3b | $1.8 \cdot 10^{-15} \Delta p$ | $1.0 \cdot 10^{-6} \Delta p$ |
| Figure 3c | $4.3 \cdot 10^{-15} \Delta p$ | $2.8 \cdot 10^{-7} \Delta p$ |

Consider the layered model (Fig. 3a). We observe that high pressure drops ($\Delta p > 10^8$ Pa) are needed to get results different from the CL solution. First at $\Delta p \geq 10^{12}$ Pa are the results equal to the VL solution. This is in accordance with $\mathcal{N}$ reported in Table 2, indicating strong capillary forces. If we use periodic BCs and set the initial saturation distribution equal to the CL distribution, we will have no vertical redistribution of the fluids, and hence all streamlines will be parallel to the layers. Thus, there will be no mixing of fluids between the layers independently of the magnitude of the pressure drop. The consequence is that the steady-state saturation distribution is equal to the CL distribution for all pressure drops. Hence, the upscaled relative permeability will be rate independent. This is verified by numerical computations. The resulting upscaled relative permeability curves coincide with the CL solution with fixed BCs as seen in Fig. 3a. However, the fractional flow will be constant within each layer, and hence this solution also represents a viscous limit [recall Eq. (13)], this one in capillary equilibrium. It is important to notice that this viscous limit is *not* the same as VL (where constant fractional flow is assumed). The latter VL solution is equal to the input relative permeability curve, which again is equal to the VL solution for fixed BCs, see Fig. 3a. Thus, we have identified two viscous limit solutions that are quite different from each other. We get the same results also with other initial distributions.

Next, consider the model with a barrier (Fig. 3c). Figure 4 shows the steady-state water saturation for different pressure drops at $\tilde{S}_w = 0.70$: capillary dominant ($\Delta p = 10^5$ Pa); mixed forces ($\Delta p = 10^8$ Pa) and viscous dominant ($\Delta p = 10^{11}$ Pa). In VL, the saturation is equal in the two rocks since the relative permeability curves are equal. For $S_w > 0.5$, capillary forces distribute water into the high permeability rock. This makes the barrier less permeable for water, and thus the upscaled relative permeability is lower in CL than in VL. For intermediate pressure drops, say $\Delta p = 10^8$ Pa, capillary forces distribute water into the high permeability rock so that water flows easier there. At the same time, the pressure drop is high enough so that water flows into the low permeability rock, making the barrier more permeable for water. This is evident from Fig. 4. Overall, this results in higher upscaled water relative permeability than in VL, and explains why the convergence from CL to VL is not monotone as seen in Fig. 3c.

Upscaling of the two first models (Fig. 3a, b) correspond to arithmetic and harmonic averaging, respectively. By comparing the CL and VL results with the analytic results in Fig. 2 ($\alpha = 50\,\%$), we see that these are equal. That the general steady-state solution converges to CL and VL serves as a verification that our upscaling routines are correct.

For the purpose of judging the balance between viscous and capillary forces in the model, the capillary number should be close to 1 at the transition between CL and VL. The micro-

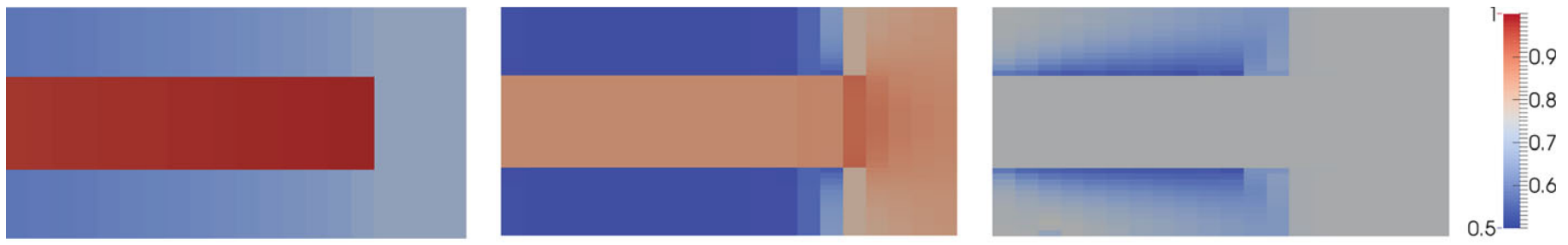

**(a)** $\Delta p = 10^5$ Pa **(b)** $\Delta p = 10^8$ Pa **(c)** $\Delta p = 10^{11}$ Pa

**Fig. 4** Saturation distribution at steady state ($\tilde{S}_w = 0.70$) for different pressure drops, $\Delta p$, for the barrier model in Fig. 3c. In CL the saturation is 0.95 and 0.58 in rock 1 and rock 2, respectively, and in VL the saturation is 0.70 in both rocks

scopic capillary number, Ca, is a few orders of magnitude smaller than 1 in the middle of the transition for the layered model (Fig. 3a), while it is off by many orders of magnitude for the two other models. Our proposed scale-dependent capillary number, $\mathcal{N}$, is about two orders of magnitude larger than 1 in the middle of the transition for the layered models (Fig. 3a, b). For the barrier model (Fig. 3c), $\mathcal{N} \sim 1$ in the middle of the transition zone. Thus, the proposed capillary number, $\mathcal{N}$, better estimates the balance of the two active forces for these models. However, we are not able to tell from $\mathcal{N}$ where the transition from CL to VL starts and ends.

## 5.2 Models Representative of Sedimentary Heterogeneity

In this section we use reservoir models similar to those considered in Nordahl et al. (2014). These models represent a key heterogeneity scale for fluvial reservoirs, since the capillary or viscous dominance is unknown. Moreover, this scale includes the most important sedimentary heterogeneity for these reservoirs Nordahl et al. (2014). Generally, such reservoirs have complex geometry and anisotropic permeability spanning several orders of magnitude.

First, we consider several small sections of dimension $100 \times 100 \times 5$ m. This is a typical size of a simulation cell, and the aim is to study steady-state upscaling on these. A wide range of sections from different sectors of the field has been examined. We therefore believe that the results and conclusions drawn are representative for this class of reservoirs. However, due to lack of space, we only present detailed results from one such section.

In the second part we consider a larger model, to represent flow within a reservoir zone. This model is large enough to represent flow between an injector and producer well pair, but still small enough to run fine-scale simulations. The aim is to study how well a coarser upscaled model captures the flow pattern and production data, with a special focus on different steady-state approaches.

### *5.2.1 REV Analysis*

Local upscaling methods like the ones we study in this work are best understood by the concept of asymptotic homogenization. Multiphase flow is governed by partial differential equations, which we then seek to homogenize over a scale of heterogeneity for the material. The heterogeneity of the material gives rise to rapidly oscillating coefficients for the equations. Central to homogenization is the concept of a representative elementary volume (or element), denoted REV. In our case, an REV is the smallest volume over which a measurement can be made that is representative of the reservoir element of interest. A key reference for the REV concept in our setting is found in Bear (1972), while Nordahl et al. (2014) contains a more recent presentation. Still, we recapture here that for the REV concept to be applicable, we depend on separation of scales. That is, being able to identify volumes of investigation

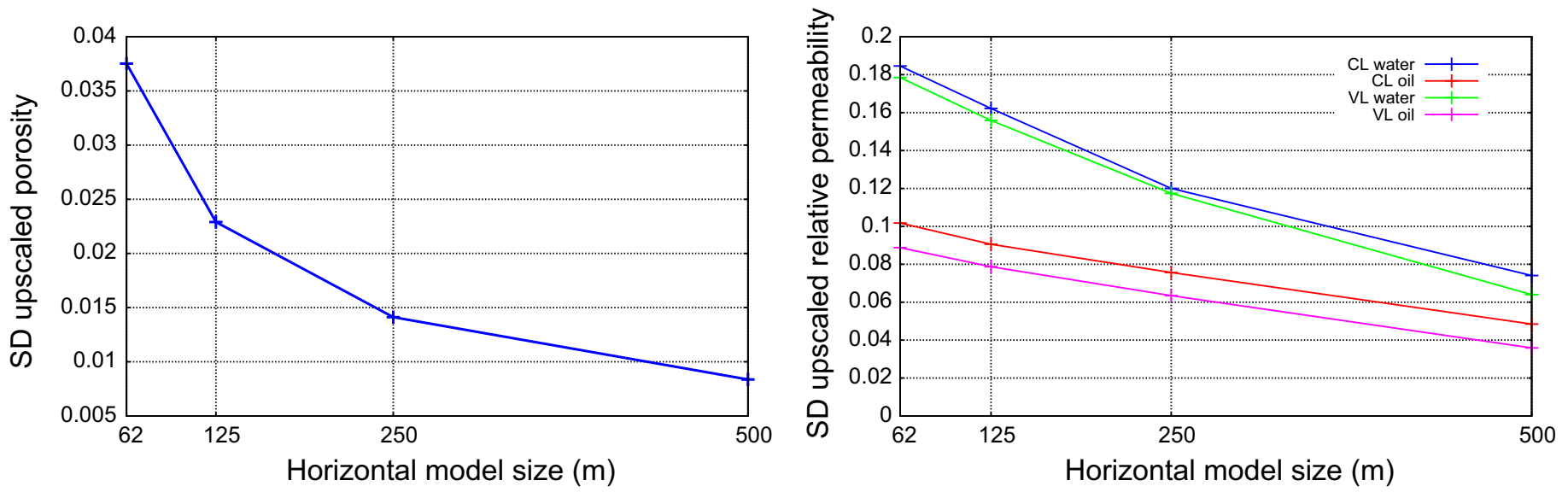

**(a)** Upscaled porosity **(b)** Upscaled relative permeability

**Fig. 5** Standard deviation (SD) for upscaled porosity (**a**) and upscaled relative permeability (**b**) plotted against horizontal model size for a representative sector model

where measured reservoir properties vary slowly locally. In other words, identify scales where homogenization makes sense.

For the models considered in this work, we have performed an REV analysis to identify REVs for different properties. This was done by cropping the original sector model into disjoint submodels of a given size. Then each submodel was upscaled and the mean and the standard deviation (SD) of the upscaled values were calculated. This was done for different sizes of the submodels, but the ratio between the horizontal and the vertical size was kept constant at 40. Porosity, permeability and relative permeability in the two limits were considered. Porosity was upscaled with volume weighted averaging and absolute permeability with a pressure solver as explained in Sect. 3. For relative permeability we used the $L^2$-norm to measure the difference between two relative permeability curves, and the mean curve was interpreted as a pointwise mean. The results for porosity and relative permeability for one sector model are shown in Fig. 5.

The target cell size for simulation scale is in our case approximately $100 \times 100 \times 5$ m. Consequently, we can only expect porosity to have a separation of scales here. Permeability is not too far off, but for relative permeability we are not at a representative volume for our simulation cell size. Moreover, relative permeability is probably not at an approximate REV even within the total size of our models (i.e., a potential REV is probably at the kilometer scale here). Hence, we chose to upscale each coarse scale cell separately for our simulation models, rather than attempting rock typing with a selected set of relative permeability curves.

### *5.2.2 Steady-State Upscaling*

Consider the section model displayed in Fig. 6, where the horizontal permeability distribution is shown together with the input relative permeability and capillary pressure curves for the 15 different anisotropic rock types. The model contains complex geometry and the permeability spans more than 4 orders of magnitude. Each rock type has individual flow properties that are not produced from some master curve, such as J-function scaling. We see that the model is periodic in the $x$-direction. This is because the original model has been mirrored. Thus, we only consider flow in the $x$-direction. In Table 3 different pressure drops are related to the capillary numbers and typical flow rates.

The test procedure is the same as for the synthetic layered model. We perform steady-state upscaling in VL and CL, and with the general steady-state method for different pressure drops. Both fixed and periodic BCs are used. The results are given in Fig. 7.

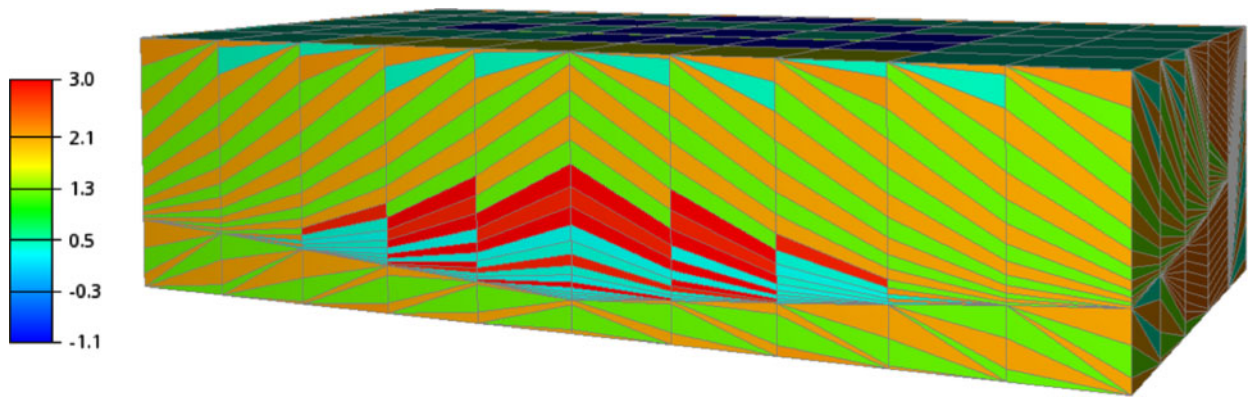

**(a)** Horizontal permeability distribution (mD) on logarithmic scale

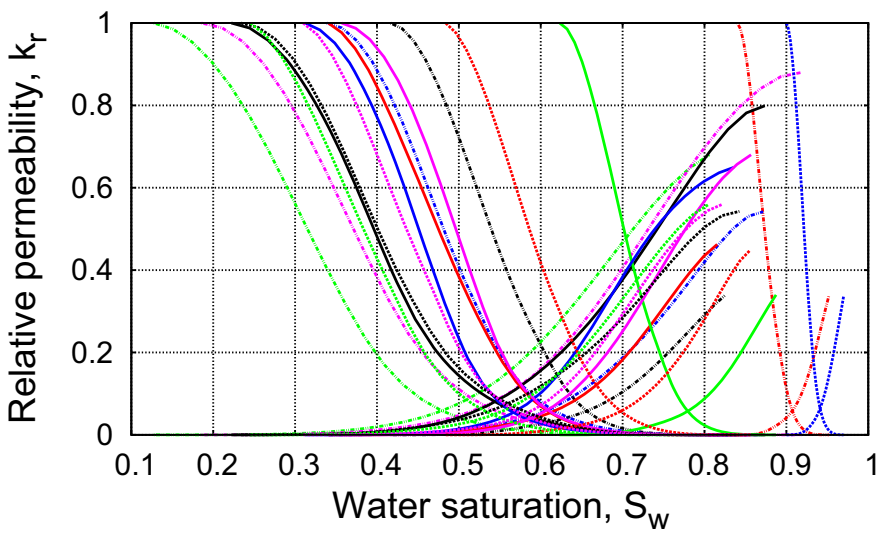

**(b)** Input relative permeability curves

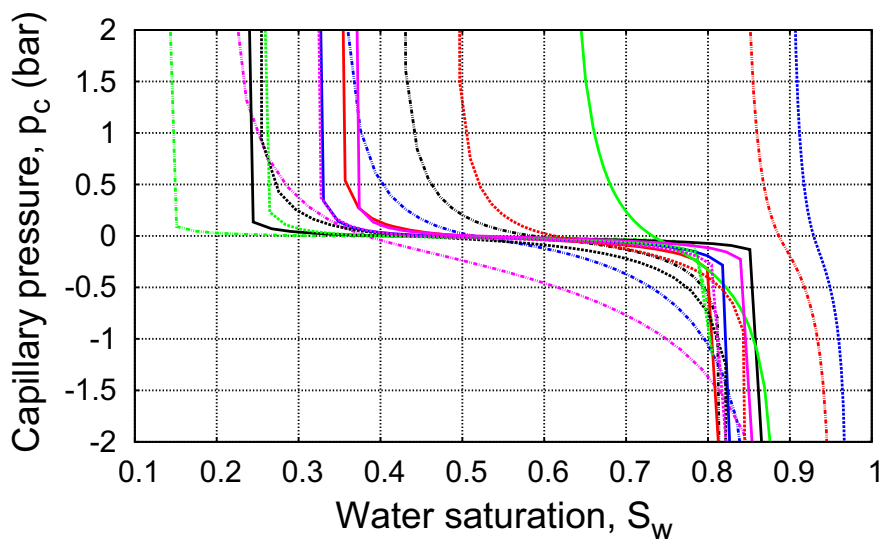

**(c)** Input capillary pressure curves

**Fig. 6** Realistic section model from a fluvial reservoir on the NCS represented in a corner-point grid. The original model is mirrored in the $x$-direction to create a periodic model. The dimension of the mirrored model is $200 \times 100 \times 5$ m. Each of the 15 rock types have uniquely generated flow properties. All curves are cut at $\pm 2$ *bar* to better see the curvature

**Table 3** Pressure drop, $\Delta p$, related to the microscopic capillary number, Ca (Eq. (15)), interpreted as $\text{Ca} = \frac{\mu u}{\sigma} = \frac{\tilde{k}\Delta p}{\sigma L}$, the macroscopic capillary number, $\mathcal{N}$ (Eq. (17)) and interstitial (pore) velocities, $v_p = u_p/\phi$ for the realistic model in Fig. 6. We use $k_r = 0.5$ and otherwise upscaled quantities to calculate $v_p$

| $\Delta p$ (Pa) | $\Delta p$ (bar) | Ca | $\mathcal{N}$ | $v_o$ (ft/day) | $v_w$ (ft/day) |
|---|---|---|---|---|---|
| $10^2$ | $10^{-3}$ | $5.5 \cdot 10^{-12}$ | $5.2 \cdot 10^{-5}$ | $2.1 \cdot 10^{-5}$ | $1.1 \cdot 10^{-4}$ |
| $10^4$ | $10^{-1}$ | $5.5 \cdot 10^{-10}$ | $5.2 \cdot 10^{-3}$ | $2.1 \cdot 10^{-3}$ | $1.1 \cdot 10^{-2}$ |
| $10^5$ | $10^0$ | $5.5 \cdot 10^{-9}$ | $5.2 \cdot 10^{-2}$ | $2.1 \cdot 10^{-2}$ | $1.1 \cdot 10^{-1}$ |
| $10^6$ | $10^1$ | $5.5 \cdot 10^{-8}$ | $5.2 \cdot 10^{-1}$ | $2.1 \cdot 10^{-1}$ | $1.1 \cdot 10^0$ |
| $10^8$ | $10^3$ | $5.5 \cdot 10^{-6}$ | $5.2 \cdot 10^1$ | $2.1 \cdot 10^1$ | $1.1 \cdot 10^2$ |

Consider the results with fixed BCs (Fig. 7a). We see that for small flow rates general steady-state upscaling coincides with CL and that the upscaled curves converge monotonically to VL as the pressure drop increases. The results with periodic BCs, see Fig. 7b, have the same behavior, and the results are very much equal. Observe that the CL and VL solutions are equal whether we use fixed or periodic BCs. The upscaled fractional flow (Fig. 7c) does also converge monotonically from CL to VL.

In Fig. 8 we see the fractional flow of water at steady state for three different pressure drops when periodic BCs are used. For a small pressure drop ($\Delta p = 1$ Pa), the system is capillary dominated, and we see that the fractional flow is nonconstant. As we increase the pressure drop, and thus the system becomes more viscous dominated, the fractional flow becomes nearly constant at $\Delta p = 10^8$ Pa. Hence, we have demonstrated that we converge

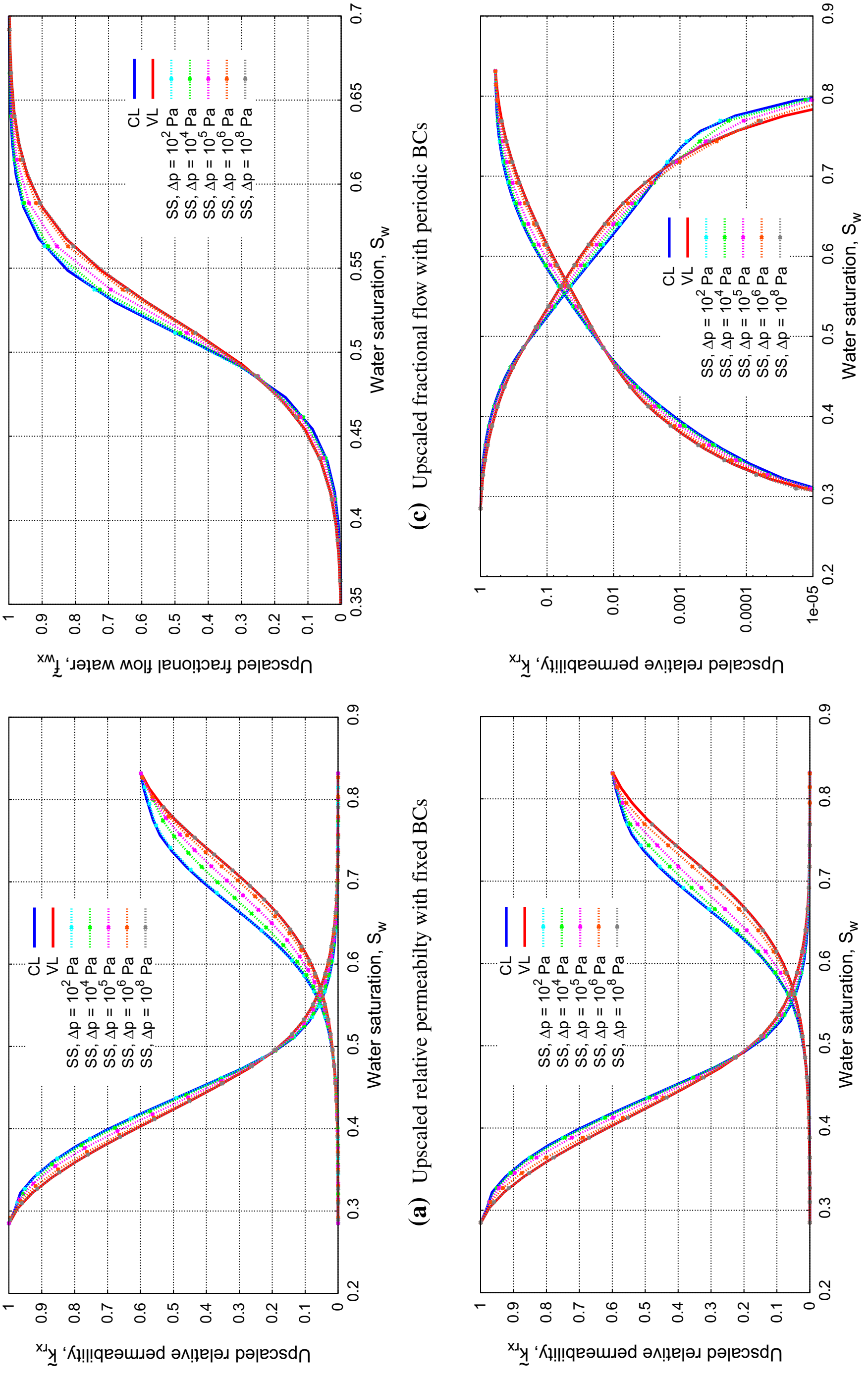

**(a)** Upscaled relative permeabilty with fixed BCs

**(b)** Upscaled relative permeabilty with periodic BCs

**(c)** Upscaled fractional flow with periodic BCs

**(d)** Logarithmic plot of upscaled relative permeability with periodic BCs

**Fig. 7** Results from steady-state upscaling (SS) of the realistic model in Fig. 6 for different pressure drops, $\Delta p$, and in CL and VL. The pressure drop direction is in the $x$-direction

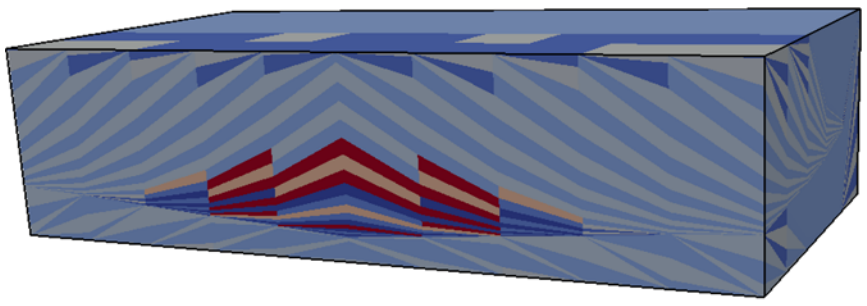

**(a)** Capillary dominated, $\Delta p = 1$ Pa ($\tilde{f}_w = 0.39$)

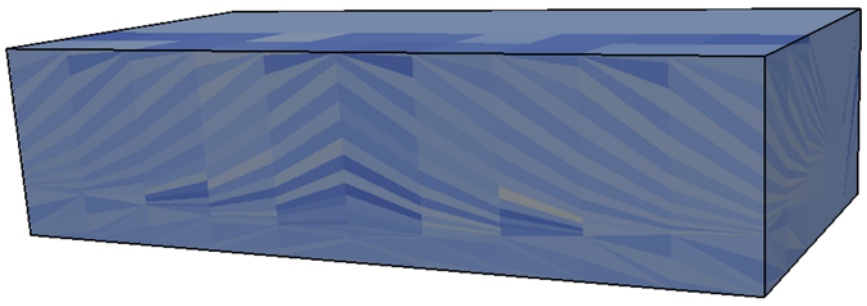

**(b)** Mixed forces, $\Delta p = 10^5$ Pa ($\tilde{f}_w = 0.37$)

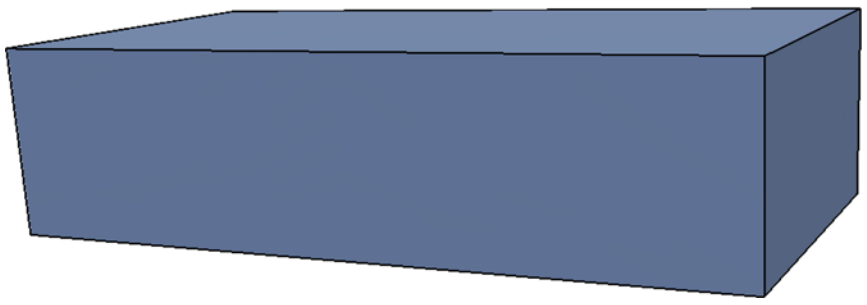

**(c)** Viscous dominated, $\Delta p = 10^8$ Pa ($\tilde{f}_w = 0.36$)

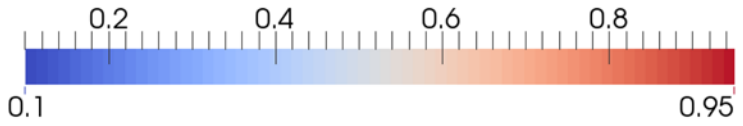

**Fig. 8** Fractional flow of water at steady state for different pressure drops, $\Delta p$, for the realistic model (Fig. 6). Periodic BCs are used and the average saturation is 0.5

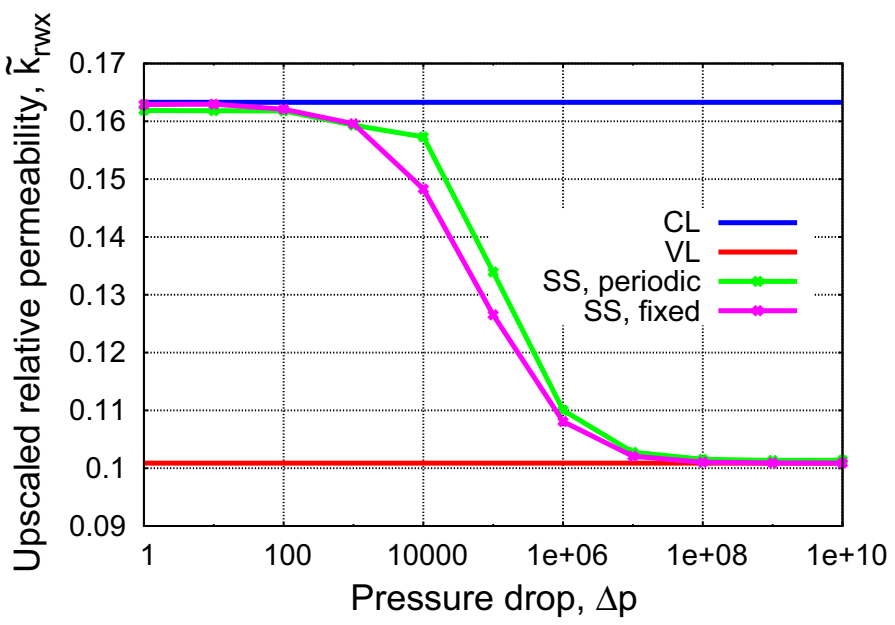

**(a)** $S_w = 0.61$

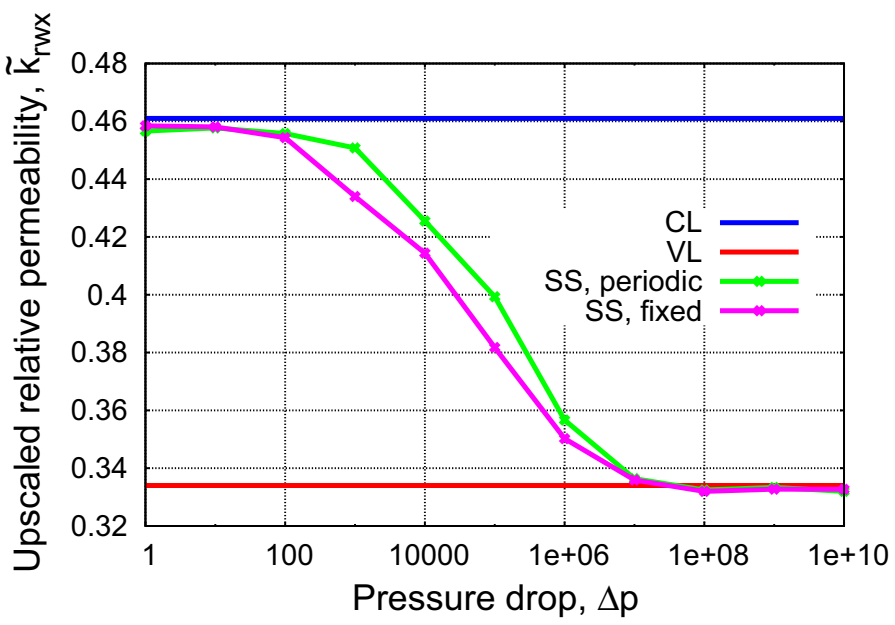

**(b)** $S_w = 0.72$

**Fig. 9** Upscaled relative permeability of water, $\tilde{k}_{rwx}$, is plotted against the pressure drop, $\Delta p$, for both fixed and periodic BCs, and at two different saturations, $S_w = 0.61$ (**a**) and $S_w = 0.72$ (**b**)

to VL as the pressure drop increases, even though we have not specified constant fractional flow on the inlet.

The difference between the results when using fixed and periodic BCs is illustrated in Fig. 9, where the upscaled relative permeability is plotted against the pressure drop at two different water saturations. We see that the differences are relatively small and that the transition from CL to VL occurs over approximately five orders of magnitude of the pressure drop. The results start to move away from CL at $\Delta p \approx 100$ Pa, which corresponds to very low water flow rates around $10^{-4}$ ft/day, and meet VL at $\Delta p \approx 10^7$ Pa, or at water flow rates around 11 ft/day, which is about one order of magnitude larger than a typical reservoir flow rate. Furthermore, we see from Table 3 that $\mathcal{N} \sim 1$ inside the rate-dependent interval. This is also the case for most of the other sections we have considered. However, the values of $\mathcal{N}$ at the start- and end-points of the rate-dependent region are not the same for all models. Similar conclusions apply to the oil curves. It is evident from Table 3 that Ca fails to model the transition from CL to VL with many orders of magnitude.

Recall that the macroscopic capillary number, $\mathcal{N}$ (Eq. (17)), is defined by taking an integral average over a range of upscaled water saturations. In Fig. 10 we have plotted $\mathcal{N}$ against different choices of $\epsilon$, which is the fraction of the saturation range that is excluded at each end of the integration interval. We conclude that $\mathcal{N}$ is relatively stable with respect to $\epsilon$ as long as $\epsilon > 0.05$.

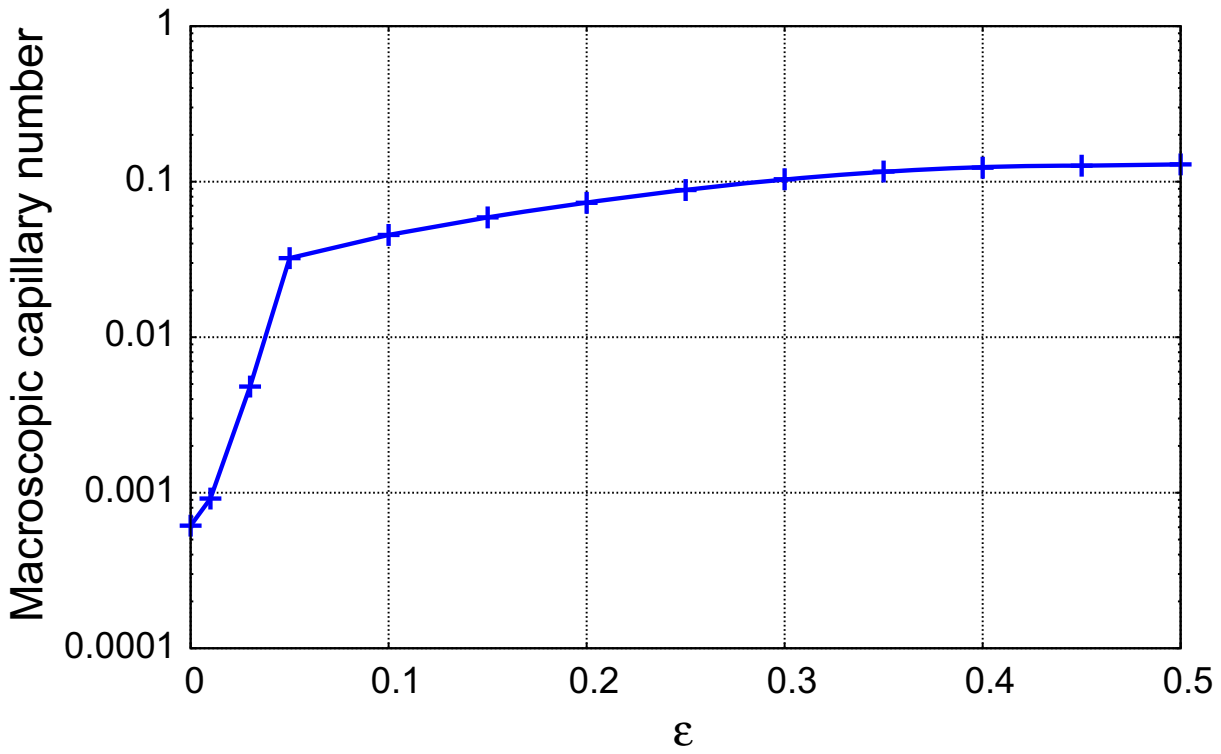

**Fig. 10** Plot of the macroscopic capillary number, $\mathcal{N}$, on a logarithmic scale, against $\epsilon$ as defined in Sect. 4 for the realistic model in Fig. 6 with a pressure drop of $10^5$ Pa

### *5.2.3 Reservoir Simulation*

We now examine the different upscaling approaches on dynamic flow scenarios on a larger field model. A fluvial reservoir model from the NCS is represented on a regular Cartesian grid with 400,000 cells, see Fig. 11. The model is originally represented on a corner-point grid, but we have regularized it to make the simulations more efficient. The reservoir dimension is $2000 \times 1000 \times 20$ m, and the discretization levels are 100, 50 and 80 in the $x$-, $y$- and $z$-directions, respectively. The model has seven different rock types, each with its unique relative permeability curve and capillary pressure curve, while porosity and permeability may vary within a rock type. The fluid data are listed in Table 1. Two vertical wells, one rate-controlled injector and one bottom hole pressure-controlled producer, are placed 700 m apart. They are placed away from the boundary, where no-flow conditions are imposed, to minimize boundary effects. The wells percolate the whole model in the $z$-direction and are completed in all layers. We initialize the reservoir with $S_w = S_{wir}$, that is the irreducible water saturation, and $p = 200$ bar. The bottom hole pressure of the producer is kept constant at 200 bar. The injection rate is also kept constant within each simulation, but a series of scenarios with different rates are run. The start date for our simulations is October 1, 2010. As in the upscaling procedures, we disregard gravity, so that we only consider capillary and viscous forces. We use OPM for upscaling and the commercial software ECLIPSE 100[3] for field simulations. For the ECLIPSE runs we use the fully implicit black oil formulation with the Peaceman well model (Peaceman 1983).

The test setup is as follows. First, a fine grid simulation is performed on the original model. This is used as a reference solution. Then, we create a coarse grid and populate it with upscaled quantities. For permeability we use flow-based anisotropic local upscaling, while for relative permeability we use the steady-state upscaling techniques with periodic BCs, though restricted to the CL and VL approaches. In this way, each coarse grid cell has its unique (upscaled) relative permeability curve. We only consider isotropic relative permeability in the coarse model, that is, we use the upscaled relative permeability originating from flow in the $x$-direction. For other parameters, like porosity and irreducible water saturation, we

[3] Reservoir simulation software from Schlumberger™.

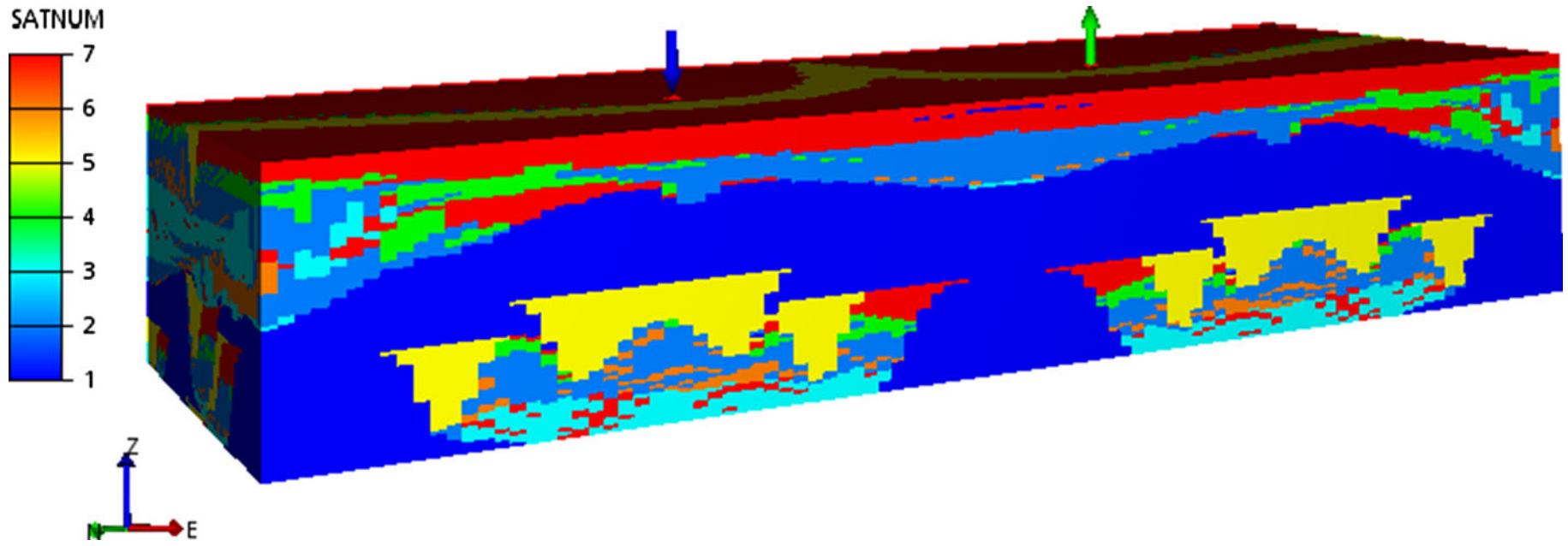

**Fig. 11** Representative fluvial reservoir model taken from a field on the NCS. The figure displays the rock type (SATNUM) distribution and the position of the wells. The dimension of the model is 2000 × 1000 × 20 m, and it is represented in a regular Cartesian grid with a total of 400,000 cells, each of dimension 20 × 20 × 0.25 m. The *horizontal* permeability variation for the plane intersecting the wells is displayed in Fig. 12

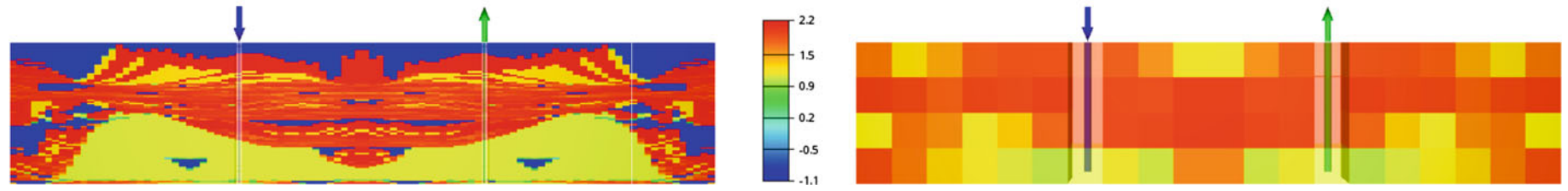

**(a)** Fine model (FM) **(b)** Coarse model (CM)

**Fig. 12** Horizontal permeability (logarithm) of (**a**) the fine model (FM), and (**b**) the upscaled coarse model (CMR) for the intersection parallel to the wells

use volume weighted averaging. The discretization levels in the coarse grid are 20, 10 and 4 in the $x$-, $y$- and $z$-directions, respectively. Thus, there is a total of 800 cells, each of size 100 × 100 × 5 m, which is the typical size of a simulation cell. Before upscaling, each coarse cell is mirrored in the $x$-direction. The wells in the coarse grid are placed at the same location as in the fine grid. Hereafter, we denote the coarse model by CM and the fine model by FM. Figure 12 shows profile views of the horizontal permeability for these two models. With this setup one would expect numerical dispersion due to the coarser discretization. To overcome this, we have also created a fine grid model with parameters taken from the coarse model. We denote this model the coarse model refined (CMR). For all simulations on FM we use relative permeability curves and capillary pressure curves taken from the original field model. For CM and CMR we use relative permeability upscaled in either CL or VL, and denote these realizations by CM CL, CM VL, CMR CL and CMR VL, respectively. In the VL cases, we disregard capillary forces by setting $p_c = 0$, while we use the upscaled capillary pressure curves produced by CL upscaling for the CL cases.

We focus our attention on two different injection rates, 20 and 200 $m^3/day$ (standard cubic meter per day). The corresponding differences in well pressures at water breakthrough for the different models are listed in Table 4. An injection rate of 200 $m^3/day$ corresponds to an average water front velocity of ~1.2 ft/day, which is quite typical. Furthermore, the average pressure drop over each coarse cell is about $1.4 \cdot 10^6$ Pa. Figure 13 shows a scatter plot of the macroscopic capillary number, $\mathcal{N}$, for all coarse cells with $\Delta p = 1.4 \cdot 10^6$ Pa. For the majority of the coarse cells, $\mathcal{N}$ is in the range [1, 4]. We can neither neglect capillary nor viscous forces based on these results, but they indicate stronger viscous forces than capillary forces. For an injection rate of 20 $m^3/day$, $\mathcal{N}$ is about ten times lower, so we expect relatively stronger capillary forces.

**Table 4** Difference in well pressure (in bar) at water breakthrough for different models and at two different injection rates

| | FM | CMR CL | CMR VL | CM CL | CM VL |
|---|---|---|---|---|---|
| 20 $m^3$/day | 9.5 | 10.0 | 11.0 | 11.5 | 13.0 |
| 200 $m^3$/day | 101.0 | 99.0 | 111.0 | 114.0 | 128.0 |

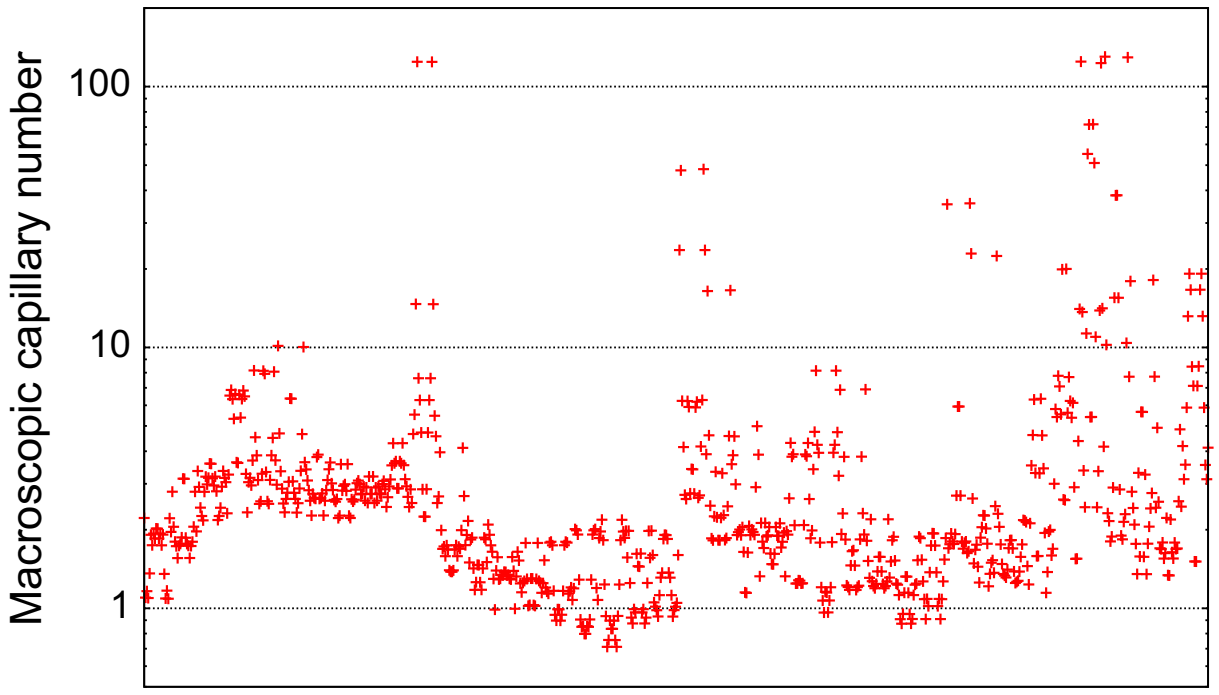

**Fig. 13** Scatter plot of the macroscopic capillary numbers $\mathcal{N}$ for all 800 coarse cells in the simulation model. A pressure drop of 14 bar $= 1.4 \times 10^6$ Pa is used over each cell

Figure 14 displays two coarse cells and their corresponding upscaled fractional flow curves. The results in Fig. 14a illustrate typical upscaled curves for the cells in the flooded region between the wells. This indicates rate sensitivity in individual coarse cells. Note that there are also cells where the rate sensitivity is less, typically in nearly homogeneous cells. Figure 14b is an example of a representative submodel where the convergence from CL to VL is not monotone. This model contains high permeability rock that only barely percolates. Compared to the results for the synthetic model with a barrier (Fig. 3c), this might explain the nonmonotonic convergence.

Before moving to the actual simulations, we make a comment on the rate dependency of the solution. Since we have an incompressible system, and if we disregard capillary forces, the water cut as function of injected water is independent on the injection rate. Specially, this means that if we use the VL approach, then all results will coincide. This motivates why we choose to plot the water cut against water injected. After all, in an incompressible system, water injected is simply time scaled by the rate.

In Fig. 15 the water cut at the producer is plotted for different models. We have also included the FM case with $p_c = 0$, representing the viscous limit scenario. When the injection rate is 200 $m^3$/day (Fig. 15a), the FM result is close to the viscous limit (FM, $p_c = 0$). Simulations (not shown here) indicate that capillary forces are negligible for flow rates above 50 $m^3$/day for FM. For an injection rate of 200 $m^3$/day the net effect of relative permeability upscaling is small, especially for the CMR cases. The difference between CL and VL upscaling is bigger when the injection rate is 20 $m^3$/day. This is due to increased capillary forces in the model. The difference between CM and CMR is small for the VL cases. For CL upscaling the difference is bigger, and CMR gives a better match with the reference solution. This might imply that for our model the upgridding effect is more prominent in CL upscaling. For all cases, we get earlier water breakthrough for CM compared to CMR due to numerical dispersion.

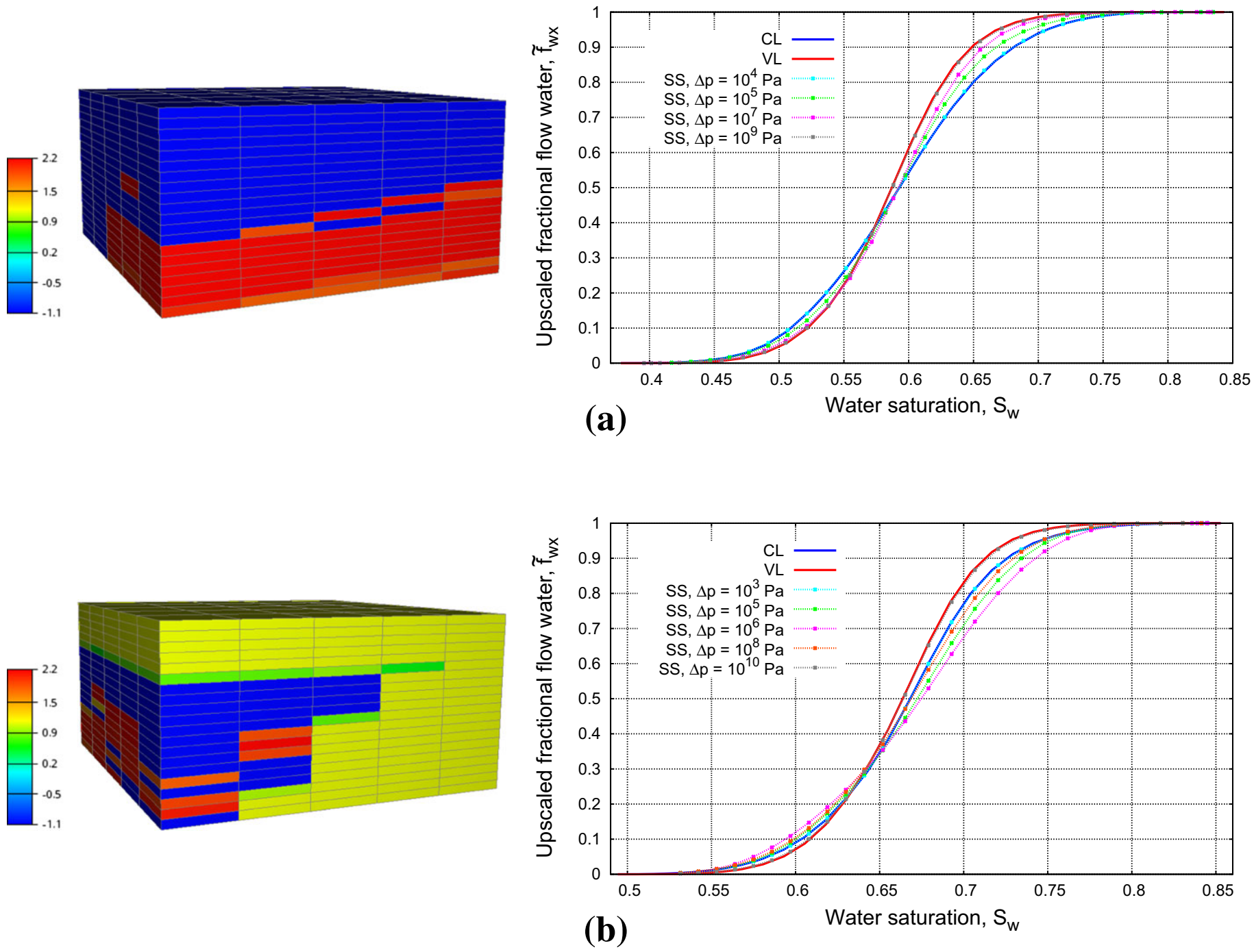

**Fig. 14** Two coarse cells from the simulation model (Fig. 12) and their upscaled fractional flow function, $\tilde{f}_w$, in CL and VL, and for rate-dependent flow (SS). The coloring in the left figures displays the horizontal permeability on a logarithmic scale. The cells belong to the flooded interwell region and the global flow direction is from left to right

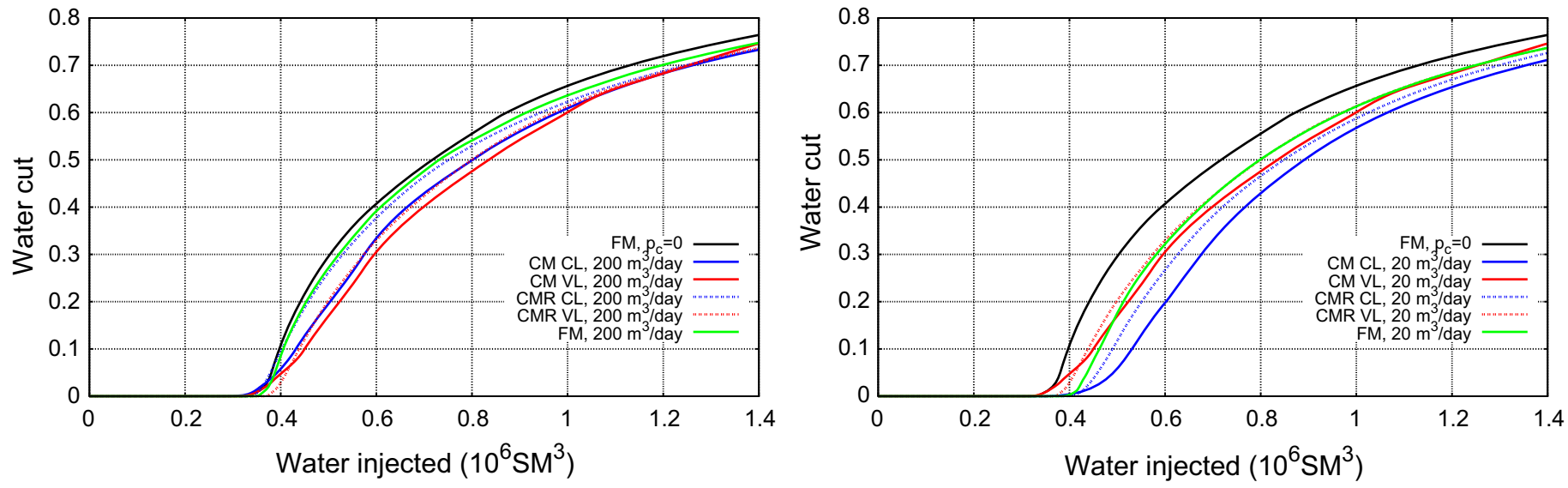

**(a)** 200 m$^3$/day **(b)** 20 m$^3$/day

**Fig. 15** Water cut at producer as function water injected when the injection rate is 200 m$^3$/day (**a**), and 20 m$^3$/day (**b**). Results for FM, CM and CMR are shown. On the two latter we use upscaled data both in CL and in VL. A FM case with $p_c = 0$ is also included

If we employ the theory of Welge (1952) for one-dimensional immiscible viscous displacement on the upscaled fractional flow curves in Fig. 14, we would expect earlier water breakthrough for the VL case compared to the CL case. In the viscous dominated case (Fig. 15a) the water breakthrough is similar for all cases. This might imply that the net effect of the rate sensitivity in individual coarse grid cells is small or that the theory of Welge is not appropriate for this scenario.

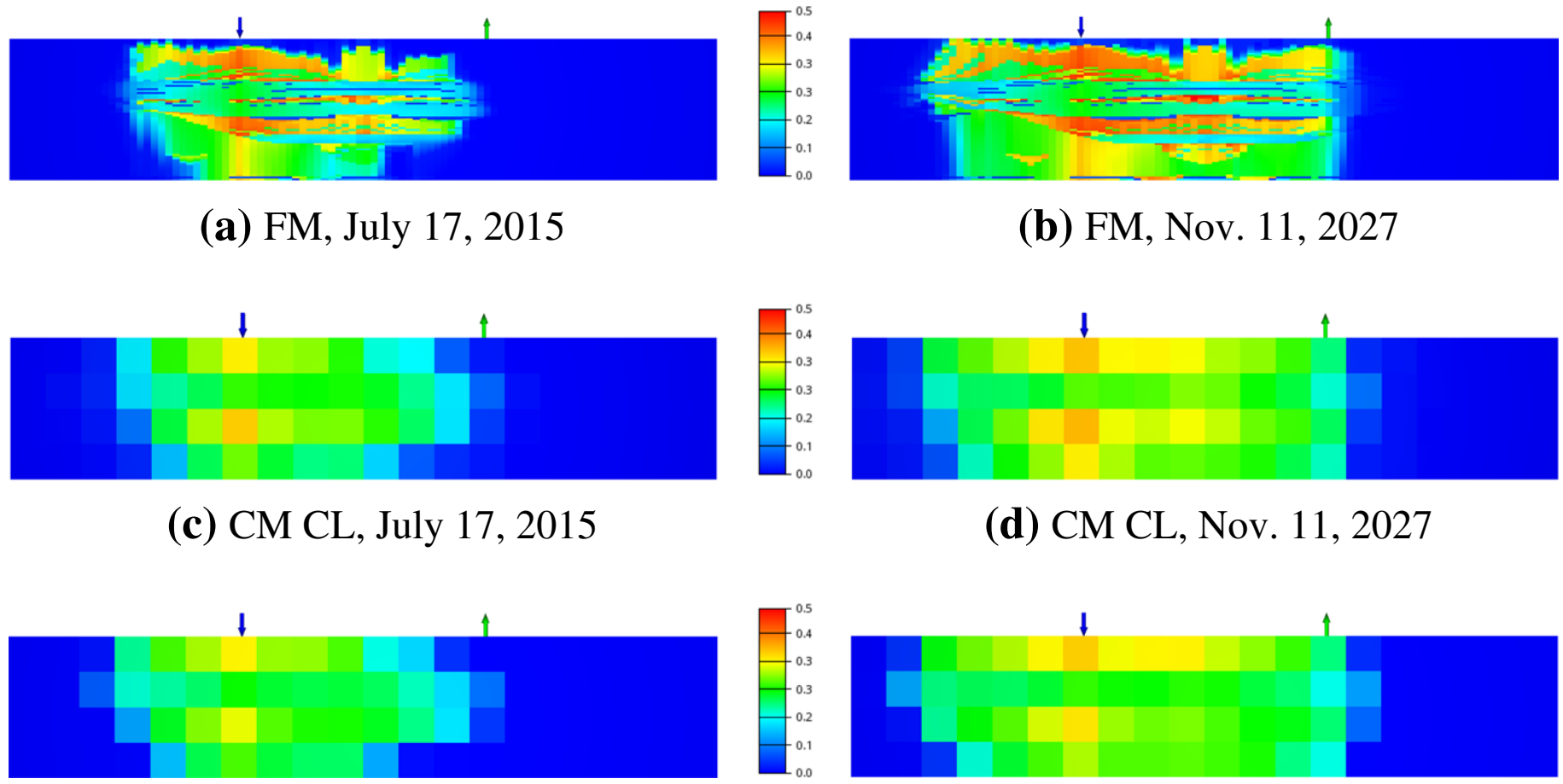

**(a)** FM, July 17, 2015 **(b)** FM, Nov. 11, 2027

**(c)** CM CL, July 17, 2015 **(d)** CM CL, Nov. 11, 2027

**(e)** CM VL, July 17, 2015 **(f)** CM VL, Nov. 11, 2027

**Fig. 16** Difference in water saturation, $S_w(\mathbf{x}, t) - S_w(\mathbf{x}, 0)$, for three simulation cases at two different times and with injection rates equal to 200 $m^3$/day. The start date for the simulations is October 1, 2010. The results are displayed in the plane parallel to the wells

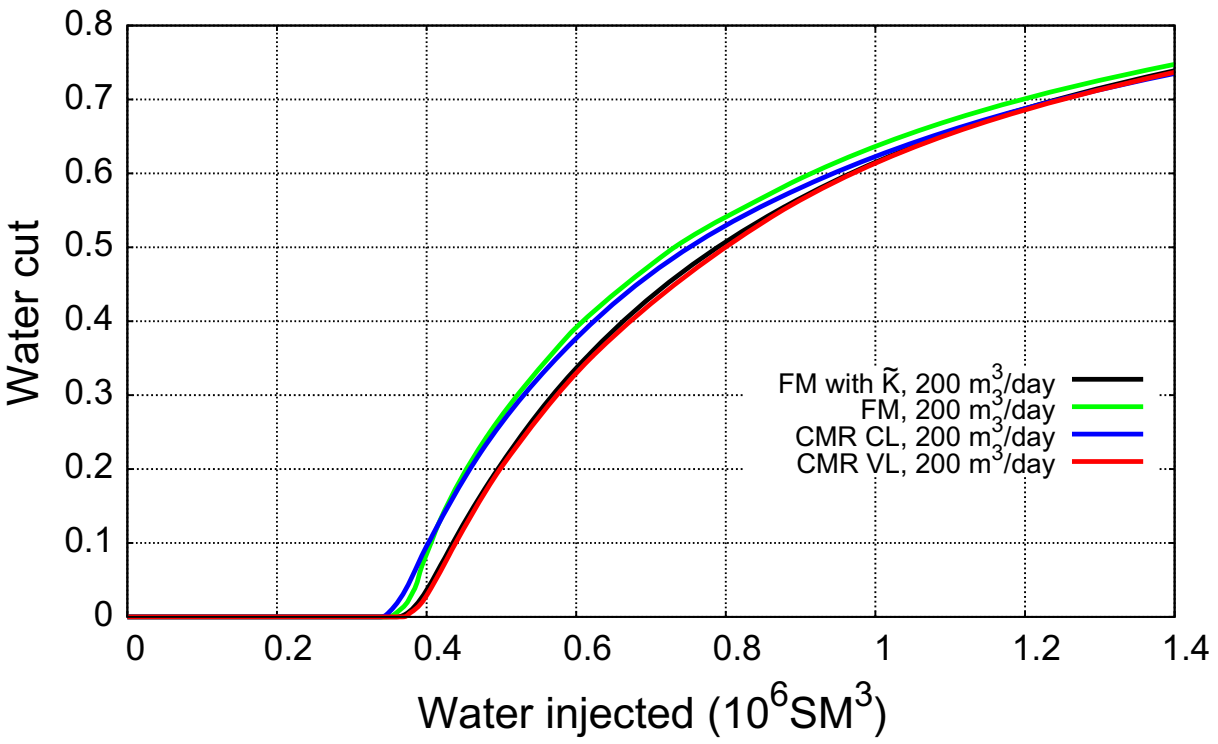

**Fig. 17** Water cut at producer as function of water injected when the injection rate is 200 $m^3$/day. Here we compare FM, CMR CL and CMR VL with the results from a fine model simulation where only permeability, porosity and critical saturations are upscaled (denoted FM with $\tilde{K}$)

Figure 16 shows, at two different times, the difference in water saturation, $S_w(\mathbf{x}, t) - S_w(\mathbf{x}, 0)$, for the three models FM, CM CL and CM VL. The coarse models are not able to capture all fine-scale variations, but the water front and the main characteristics are fairly close. There are only minor differences between the CL and VL cases visible from this figure.

To study the relative importance of permeability upscaling compared to relative permeability upscaling, the fine model was run with upscaled porosity, permeability and irreducible saturation, see Fig. 17 (denoted FM with $\tilde{K}$). For this case we only compare with CMR to remove upgridding effects. We see that FM with $\tilde{K}$ is almost identical to CMR VL, but differs from CMR CL. This is in accordance with the small capillary forces in the model and indicates that VL upscaling works fairly well. Furthermore, for this model it seems like CL upscaling works as a fortunate correction to the error introduced by permeability upscaling. This example also demonstrates that permeability upscaling is a significant source of error

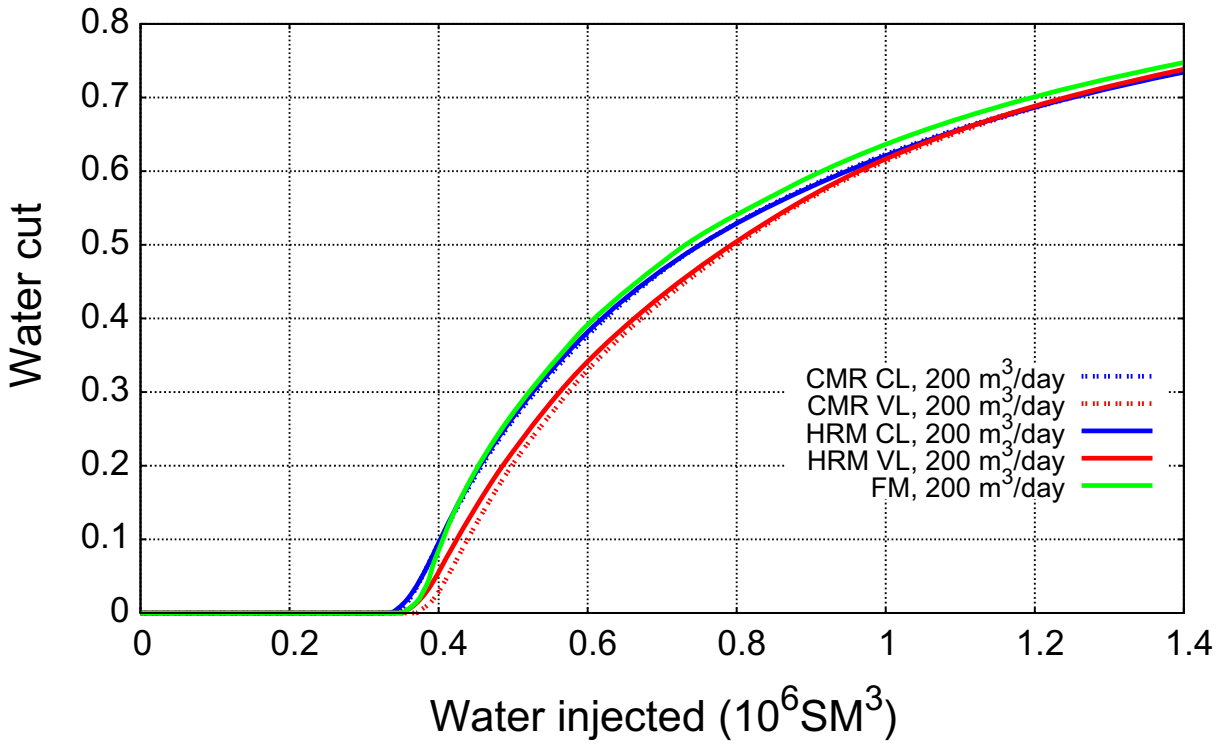

**Fig. 18** Water cut at producer as function of water injected when the injection rate is 200 $m^3$/day. Results for the fine model (FM), the coarse model refined (CMR) and the homogeneous relative permeability model (HRM) are shown. On the two latter models we use upscaled data, both in CL and in VL

in the upscaled model. However, the CL case shows that relative permeability upscaling is about equally important.

Finally, consider CMR, but now with homogeneous relative permeability calculated by upscaling the whole model as one bulk, both in CL and in VL. All parameters other than relative permeability and capillary pressure are equal to those used in CMR. We denote this model by the homogeneous relative permeability model (HRM). In Fig. 18 the water cut for these scenarios with an injection rate of 200 $m^3$/day is shown together with the results from CMR and FM. We observe that the results do not differ significantly. This example illustrates that it might be possible to use homogeneous relative permeability.

## 6 Discussion and Conclusions

Steady-state upscaling has been studied on representative three-dimensional models. The correctness of the implemented upscaling procedures was demonstrated by comparing with analytic upscaling results in the capillary and viscous limits. We have further demonstrated that the general steady-state solution converges toward the capillary limit solution as the flow rate tends to zero, and conversely, that it converges toward the viscous limit solution, defined by constancy in fractional flow, as the flow rate increases. This holds for all models we have considered, except for flow along layers with periodic boundary conditions. The convergence may fail to be monotonic for some models, in accordance with results in Virnovsky et al. (2004). The transition from the capillary to the viscous limit solution is shown to occur over several orders of magnitude of the flow rate (or equivalently pressure drop). This substantiate the need for rate-dependent steady-state upscaling in a wide range of flow scenarios.

Numerical examples demonstrate that it is of great importance whether the model contains high permeability rock with poor connectivity. If this is present, the convergence from the capillary to viscous limit may fail to be monotonic.

A *static* dimensionless capillary number that models the force balance for *transient* flow can only be approximate. Still, our proposed scale-dependent capillary number, $\mathcal{N}$, gives values close to 1 in the transition from capillary to viscous dominated flow for a wide range of models. It is thus much more suitable for this purpose than the traditional (microscopic) capillary number, Ca. Furthermore, it takes length scale into account, and it is robust and easy to calculate.

We have demonstrated that steady-state upscaling is dependent on the choice of boundary conditions (BCs). However, for the realistic section models considered here, the differences were modest. The flow rate is of much greater importance. In the capillary and viscous limits, and with a periodic reservoir model, the results are the same for these two sets of BCs. It should be noted that the differences probably would have been larger if we had considered fully periodic BCs, so that also the boundaries normal to the pressure drop direction are periodic. This is most prominent for models with layers not aligned with the global flow direction. Fully periodic BCs would furthermore produce a full upscaled tensor. An advantage of periodic BCs is that the upscaled saturation is the same at steady state as it was initially. Thus, it is easy to get upscaled relative permeabilities for a uniform distribution of upscaled saturations. One drawback with periodic BCs is that the flow rate can be significantly reduced if the model is not periodic. This is why we have chosen to always mirror the models.

Through the REV analysis for the fluvial reservoir models considered herein, we were able to identify an REV for porosity. This is in accordance with similar results in Nordahl et al. (2014). However, we were not able to identify any REV for relative permeability. This suggests that one should use unique upscaled relative permeabilities for each coarse cell.

Simulations on the reservoir scale illustrated that our model is viscous dominated for reasonable injection rates. This was also supported by our macroscopic capillary number. Taking into account the effect of permeability upscaling, viscous limit upscaling seems to be appropriate in this viscous dominated scenario.

**Acknowledgments** The research is funded by VISTA—a basic research program funded by Statoil, conducted in close collaboration with The Norwegian Academy of Science and Letters. The authors thank Statoil for giving permission to publish this work. The authors also want to thank Atgeirr Flø Rasmussen, Bård Skaflestad and Halvor Møll Nilsen, SINTEF ICT, and Håvard Berland, Statoil, for support and development of the upscaling routines in OPM, and Sindre Hilden, NTNU and SINTEF ICT, for helpful discussions. Finally, the authors are grateful for the thorough and constructive feedback from the Reviewers, which have helped to improve this paper.